\documentclass[journal]{IEEEtran}
\ifCLASSINFOpdf
   \usepackage[pdftex]{graphicx}
\else
\fi

\usepackage{amsmath}
\usepackage{amsfonts}
\usepackage{mathtools}
\usepackage{multirow}

\usepackage{algorithmic}
\usepackage{algorithm}
\usepackage{xcolor}

\usepackage{caption}
\usepackage{subcaption}

\usepackage{url}

\usepackage{comment}
\usepackage{booktabs}
\usepackage{bm}
\usepackage{enumitem}

\DeclareMathOperator*{\argmin}{arg\,min}

\begin{document}
\urlstyle{tt}

%
\title{Meta-TTS: Meta-Learning for Few-Shot Speaker Adaptive Text-to-Speech}
%
%
%

\author{Sung-Feng Huang, Chyi-Jiunn Lin, Da-Rong Liu, Yi-Chen Chen, Hung-yi Lee 
\thanks{S.-F. Huang, D.-R. Liu and Y. -C. Chen are with the Graduate Institute of Communication Engineering, National Taiwan University, Taipei 10617, Taiwan (e-mail: f06942045@ntu.edu.tw; f07942148@ntu.edu.tw; f06942069@ntu.edu.tw).}
\thanks{C.-J. Lin and H. Lee are with the Department of Electrical Engineering, National Taiwan University, Taipei 10617, Taiwan (e-mail: b08901060@ntu.edu.tw; hungyilee@ntu.edu.tw).}}



%



\maketitle

\begin{abstract}

Personalizing a speech synthesis system is a highly desired application, where the system can generate speech with the user's voice with rare enrolled recordings. There are two main approaches to build such a system in recent works: speaker adaptation and speaker encoding. On the one hand, speaker adaptation methods fine-tune a trained multi-speaker text-to-speech (TTS) model with few enrolled samples. However, they require at least thousands of fine-tuning steps for high-quality adaptation, making it hard to apply on devices. On the other hand, speaker encoding methods encode enrollment utterances into a speaker embedding. The trained TTS model can synthesize the user's speech conditioned on the corresponding speaker embedding. Nevertheless, the speaker encoder suffers from the generalization gap between the seen and unseen speakers.

In this paper, we propose applying a meta-learning algorithm to the speaker adaptation method. More specifically, we use Model Agnostic Meta-Learning (MAML) as the training algorithm of a multi-speaker TTS model, which aims to find a great meta-initialization to adapt the model to any few-shot speaker adaptation tasks quickly. Therefore, we can also adapt the meta-trained TTS model to unseen speakers efficiently. Our experiments compare the proposed method (Meta-TTS) with two baselines: a speaker adaptation method baseline and a speaker encoding method baseline. The evaluation results show that Meta-TTS can synthesize high speaker-similarity speech from few enrollment samples with fewer adaptation steps than the speaker adaptation baseline and outperforms the speaker encoding baseline under the same training scheme. When the speaker encoder of the baseline is pre-trained with extra 8371 speakers of data, Meta-TTS can still outperform the baseline on LibriTTS dataset and achieve comparable results on VCTK dataset.
Our code is now publicly available at \url{https://github.com/SungFeng-Huang/Meta-TTS/}, and we also provide some demo audio at \url{https://reurl.cc/k7l1nG}.

\end{abstract}


%
\IEEEpeerreviewmaketitle

\section{Introduction}
\label{sec:intro}
\IEEEPARstart{T}{ext-to-speech} (TTS) aims to synthesis speech from text, which has been widely used in our lives, such as Siri and Alexa. TTS can synthesize a natural human voice when trained with a large amount of high-quality single-speaker speech data~\cite{oord2016wavenet,wang2017tacotron,ren2020fastspeech} or synthesize a set of voices with multi-speaker corpora~\cite{gibiansky2017deep,ping2018deep,jia2018transfer,cooper2020zero}. These multi-speaker corpora contain a fixed set of speakers, and each speaker still requires a certain amount of speech data. However, when we want to customize our TTS to an arbitrary voice, it would be too costly to collect a sufficient amount of speech data with the corresponding voice. Therefore, learning the voice of an unseen speaker from only a few samples becomes an intriguing task for research. In this paper, we call it a ``voice cloning'' task.

There are two general approaches to deal with such task~\cite{arik2018neural}, speaker adaptation~\cite{arik2018neural,chen2018sample,Wang20adaptation,chen2021adaspeech,Song21M2VoC} and speaker encoding~\cite{jia2018transfer,Yaniv18Voiceloop,cooper2020zero,Choi20Attentron,Wang20bilevel,Cai20verification,Chien21M2VoC}.
The speaker encoding method builds a multi-speaker TTS architecture which consists of a speaker encoder and a TTS model. The speaker encoder could be pre-trained~\cite{jia2018transfer} or jointly trained~\cite{arik2018neural,chen2018sample} with the TTS model. In order to clone the voice of an unseen speaker, the speaker encoder extracts the speaker's embedding from a few speech samples. The trained TTS model could synthesize speech samples with the speaker's voice conditioned on the speaker embedding. However, the speaker encoder might suffer from generalization problems and adapt worse for unseen speakers.
On the other hand, the multi-speaker TTS architecture includes a speaker embedding look-up table instead of a speaker encoder for the speaker adaptation method~\cite{arik2018neural,chen2018sample,chen2021adaspeech}. The speaker embedding table is jointly trained with the TTS model. When learning an unseen speaker's voice, we would first randomly initialize an embedding for the new speaker, and then the embedding would be fine-tuned by the speech samples of the unseen speaker alone or with the TTS model.
As experimented in~\cite{arik2018neural, chen2018sample}, although the speaker adaptation approach performs better than the speaker encoding approach, it requires thousands of adaptation steps, which means more cloning time and computational resources are needed for high-quality voice cloning.

\begin{figure}[t]
    \centering
    \begin{subfigure}[]{0.6\linewidth}
        \includegraphics[width=\linewidth]{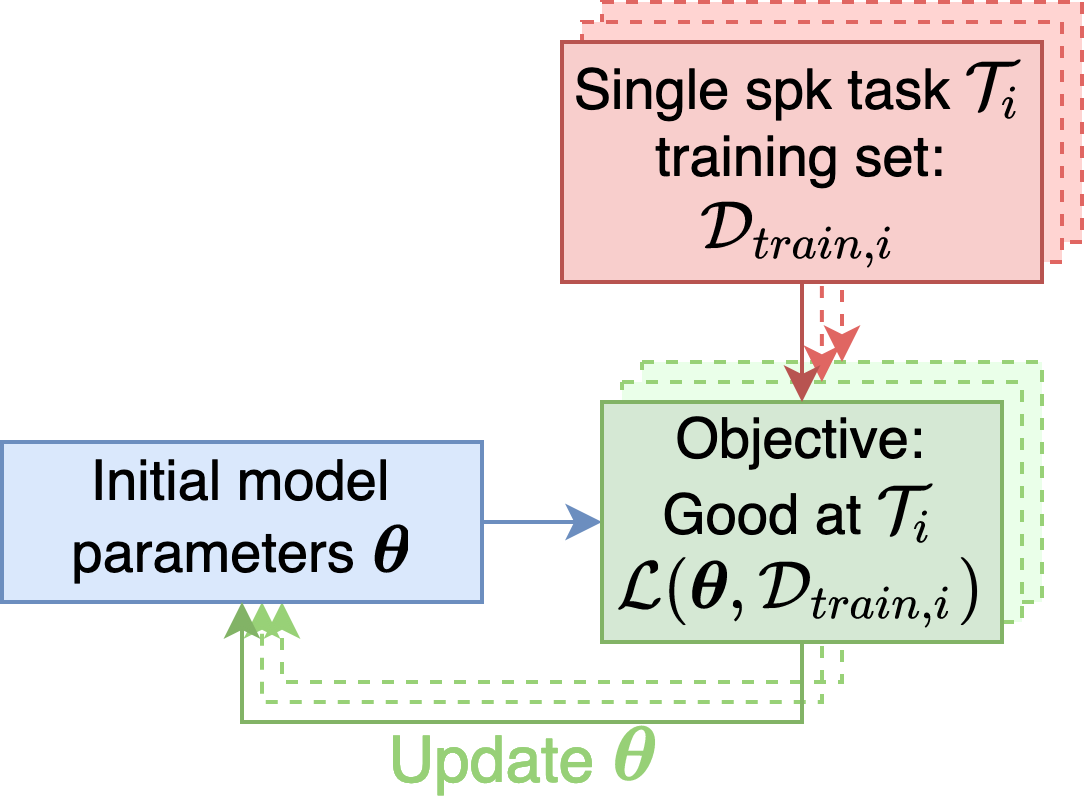}
        \caption{Multi-task learning}
        \label{subfig:multi-task-learning}
    \end{subfigure}
    \par\bigskip
    \begin{subfigure}[]{\linewidth}
        \includegraphics[width=\linewidth]{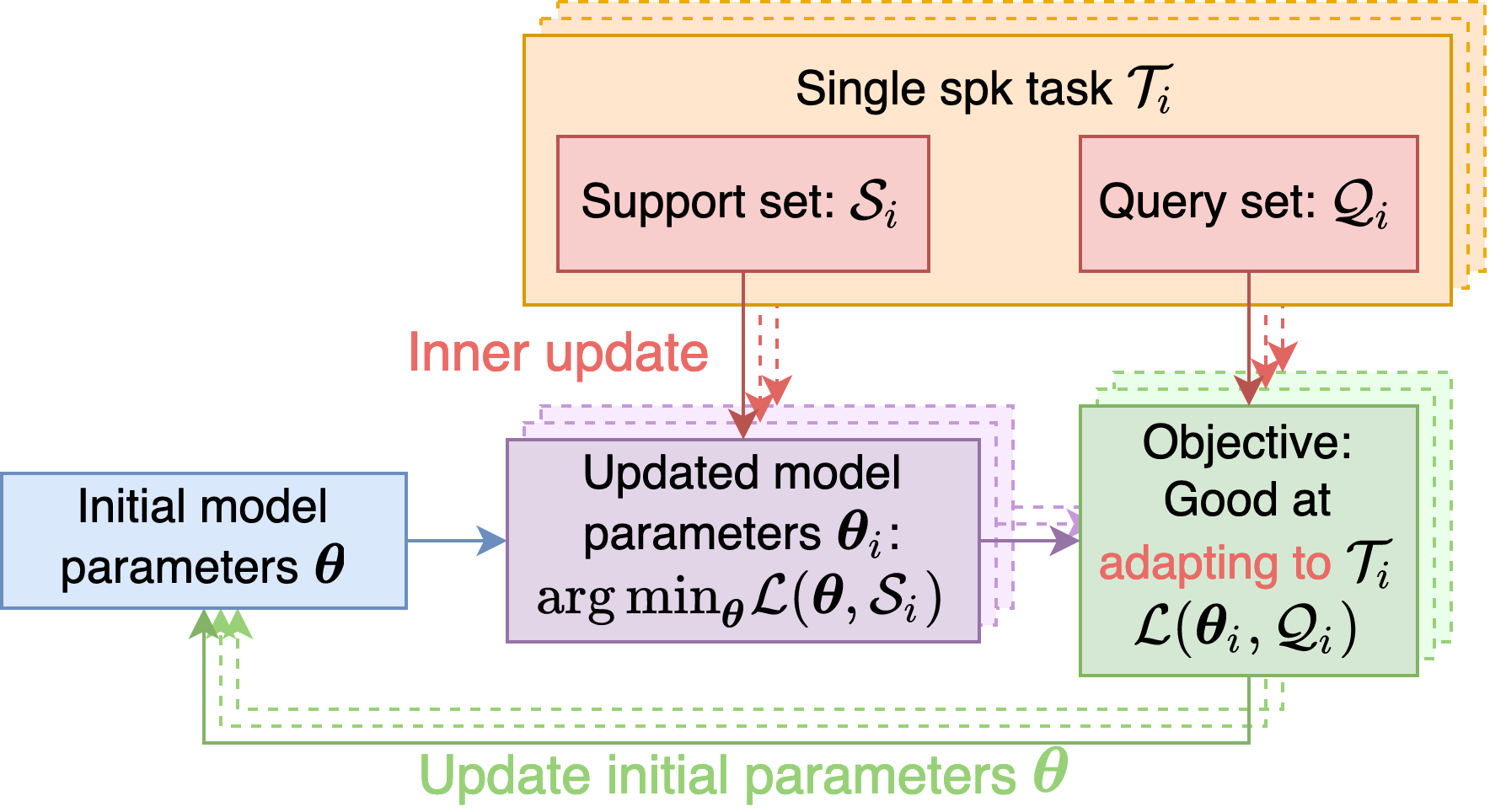}
        \caption{Meta learning}
        \label{subfig:meta-learning}
    \end{subfigure}
    \caption{Training step illustration of multi-task learning and meta learning, where ``spk'' is the abbreviation of ``speaker''.}
    \label{fig:1}
\end{figure}

To speed up the adaptation process, we propose applying meta-learning to the speaker adaptation approach in this paper. Model-Agnostic Meta-Learning (MAML)~\cite{finn2017model} is a typical meta-learning algorithm, which aims to find a parameter initialization for fast adaptation. MAML consists of two optimization loops, where the outer loop tries to find a meta-initialization, from which the inner loop can quickly adapt to new tasks with few samples. Contrary to transfer learning with a model trained by multi-task learning, since MAML forces the model to have the capability to learn new tasks efficiently, the meta-learned model would be a more suitable initialization for few-shot adaptation. The flowcharts of training with multi-task learning and meta-learning are shown in Figure~\ref{subfig:multi-task-learning} and~\ref{subfig:meta-learning} respectively.
Therefore, we propose training the multi-speaker TTS via MAML to obtain high-quality adaptation results with fewer adaptation steps. We use FastSpeech 2~\cite{ren2020fastspeech} as our TTS model architecture, which is one of the most popular single-speaker non-autoregressive TTS models.
FastSpeech 2 is further modified to a multi-speaker version as described in Section~\ref{ssec:multi-spk-fastspeech}. 
Since some modules of FastSpeech 2 are not related to the speaker identity, we do not directly apply MAML on the whole model but only to those conditioned on the speakers, where the details are in Section~\ref{sec:meta-tts}. 

In our experiments, we evaluate the setting of the 5-shot voice cloning tasks. We compare the proposed method (Meta-TTS) with the baseline of the speaker adaptation method and the baseline of the speaker encoding method. The only difference between Meta-TTS and the speaker adaptation baseline is the training algorithm (meta-training v.s. multi-task training). In contrast, the speaker encoding baseline uses an additional speaker encoder and does not require fine-tuning.
The experiment results show that Meta-TTS can synthesize high speaker-similarity speech from 5 enrollment samples with fewer adaptation steps than the speaker adaptation baseline.
Moreover, Meta-TTS outperforms the speaker encoding baseline under the same training scheme. Furthermore, when the speaker encoder of the baseline is pre-trained with extra 8371 speakers of data, Meta-TTS can still outperform the baseline on LibriTTS dataset and achieve comparable results on VCTK dataset.


\section{Problem definition}
\label{sec:problem-def}

\subsection{Tasks}
In this paper, we are dealing with 5-shot voice cloning tasks. Each task $\mathcal{T}_i$ consists of a support set (i.e. train dataset of the task) $\mathcal{S}_i=\{(\mathbf{x}_j^{\mathcal{S}_i},\mathbf{y}_j^{\mathcal{S}_i})\}_{j=1}^K$ and a query set (i.e. test dataset of the task) $\mathcal{Q}_i=\{(\mathbf{x}_j^{\mathcal{Q}_i},\mathbf{y}_j^{\mathcal{Q}_i})\}_{j=1}^Q$, where $K=5$, $Q$ is the number of examples in the query set, $\mathbf{x}_j$ indicate the inputs (speaker identity, phoneme sequence for TTS) and $\mathbf{y}_j$ represent the training targets (mel-spectrogram, duration, pitch and energy for TTS). For each task $\mathcal{T}_i$, the support examples $(\mathbf{x}^{\mathcal{S}_i},\mathbf{y}^{\mathcal{S}_i})$ and the query examples $(\mathbf{x}^{\mathcal{Q}_i},\mathbf{y}^{\mathcal{Q}_i})$ are all from the same speaker, and the model should learn the speaker's voice by fitting on the support set $\mathcal{S}_i$, then synthesize the utterance with the correct voice according to the phoneme sequence inputs of the query set $\mathbf{x}_j^{\mathcal{Q}_i}$.

\subsection{Methods}
If we directly use a randomly initialized TTS model to fit on the support set of each 5-shot task, the model could easily overfit on the support set and failed to speak the words correctly. Therefore, the related works usually use transfer learning to transfer the TTS knowledge from a trained TTS model, then only fine-tune the speaker-related modules on the downstream tasks to tune the voice. Especially, if the model is trained under multi-speaker scenario, the model could be tuned better~\cite{arik2018neural,jia2018transfer,chen2018sample,chen2021adaspeech}.
However, fine-tuning from a trained TTS model is slow (can spend from minutes to hours), and requires lots of tricks to prevent over-fitting (e.g. only tune the speaker embedding, early stopping, adaptive layer norm, etc.). So we propose using meta-learning to deal with the few-shot tasks, where the fine-tuning procedure can be compressed into under 10 seconds (or even around 1 second) to achieve high-quality results.
In Section~\ref{sec:multi-spk-tts} and~\ref{sec:spk-adaptation}, we would introduce the multi-speaker TTS and how to transfer to few-shot tasks via fine-tuning. In Section~\ref{sec:meta-tts}, we would then explain our proposed meta-learning method.

\subsection{Evaluation}
We will evaluate those trained models' adaptation capability of few-shot voice cloning tasks. First, we will generate a set of 5-shot testing tasks from the testing corpus. Then for each task, the models are required to fit on the support set and evaluate the speaker similarity and the speech quality on the query set. We will introduce the evaluation metrics in Section~\ref{sec:evaluation-metrics}.

\section{Multi-speaker TTS architecture}
\label{sec:multi-spk-tts}

In this section, we would first describe the model architecture of FastSpeech 2~\cite{ren2020fastspeech} in Section~\ref{ssec:fastspeech2}, which is originally a single-speaker TTS architecture and used as the basic structure of our model. Then we would introduce a multi-speaker version of FastSpeech 2 in Section~\ref{ssec:multi-spk-fastspeech}.

\subsection{FastSpeech 2}
\label{ssec:fastspeech2}

\begin{figure}[t]
    \centering
    \begin{subfigure}[b]{0.54\linewidth}
        \includegraphics[width=\linewidth]{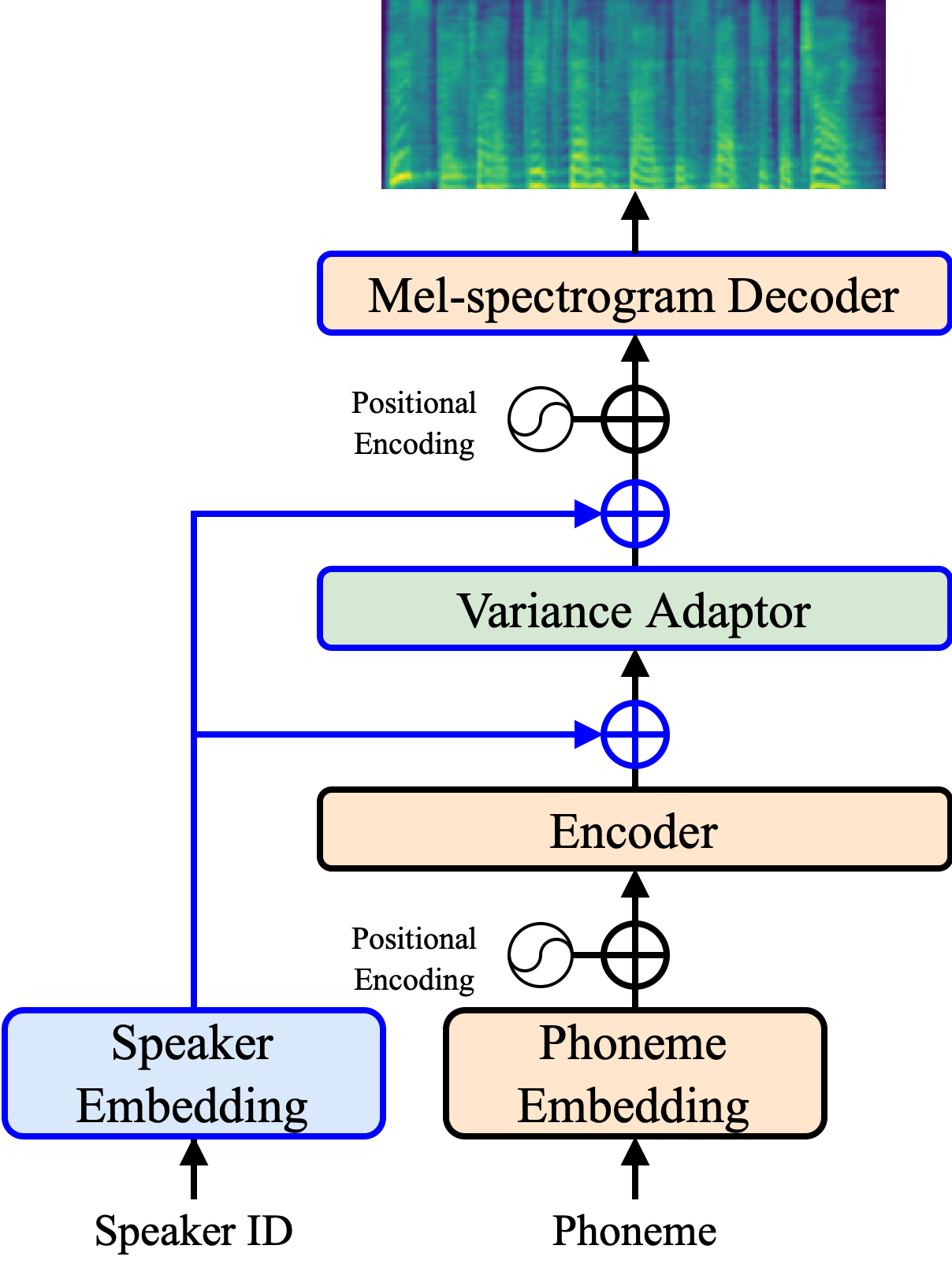}
        \caption{Multi-speaker FastSpeech 2.}
        \label{subfig:multi-spk-fastspeech}
    \end{subfigure}
    \hfill
    \begin{subfigure}[b]{0.36\linewidth}
        \includegraphics[width=\linewidth]{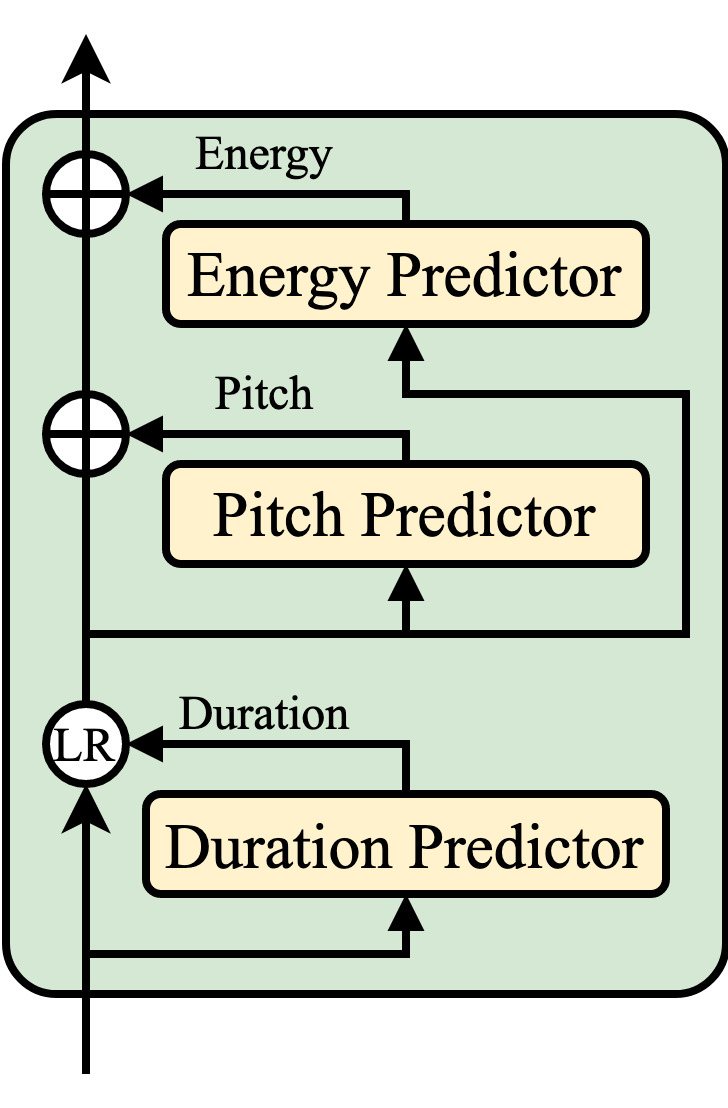}
        \caption{Variance Adaptor.}
        \label{subfig:VA}
    \end{subfigure}
    \caption{Illustration of (a) the multi-speaker TTS architecture and (b) the variance adaptor.
    The blocks with blue borders in (a) indicate that they would condition on the speaker identity.}
    \label{fig:2}
\end{figure}

Figure~\ref{subfig:multi-spk-fastspeech} shows the architecture of multi-speaker FastSpeech 2, which could reduce to the original FastSpeech 2 by removing the speaker embedding block.
FastSpeech 2 mainly consists of three modules: encoder $\bm\theta_E$, variance adapter $\bm\theta_{VA}$ and mel-spectrogram decoder $\bm\theta_D$.\footnote{We may simply use ``decoder'' to refer to mel-spectrogram decoder.}
The encoder and the decoder are stacks of transformer layers, and the variance adaptor's architecture is in Figure~\ref{subfig:VA}.
The encoder converts the phoneme embedding sequence into a phoneme hidden sequence, and then the variance adapter adds duration, pitch, and energy information into the hidden sequence. Finally, the decoder converts the adapted hidden sequence into a mel-spectrogram sequence in parallel.

\subsection{Multi-speaker FastSpeech 2}
\label{ssec:multi-spk-fastspeech}
It is well-known that speeches with the same content would not all sound the same. Furthermore, the main differences are the speaker characteristics, including speaking speed, tones, loudness, and timbre.
FastSpeech 2 decides speaking speed, tones, and loudness  (corresponding to duration, pitch, and energy predictions, respectively) by the variance adapter, and the decoder determines timbre. 
So these two modules are the significant parts that determine the speaker of the sound, while the encoder aims to extract the contextual information from the phoneme sequence.
We add the embedding of the target speaker to the inputs of both the variance adapter and the decoder so that these two modules can condition on the target speaker.
The architecture of multi-speaker FastSpeech 2 is in Figure~\ref{subfig:multi-spk-fastspeech}, where the only change to the original version is to maintain a learnable speaker embedding look-up table $E_S$.
During training, we need to train the model with a multi-speaker dataset so that $E_S$ can learn together with $\bm\theta_E$, $\bm\theta_{VA}$ and $\bm\theta_D$.

\section{Speaker adaptation via fine-tuning}
\label{sec:spk-adaptation}

Transfer learning is useful in domain adaptation, especially in few-shot cases. The model is first pre-trained in a multi-task configuration, then fine-tuned with the few labeled examples of a new task.
Since multi-speaker TTS is also a multi-task version of single-speaker TTS, we can adapt the pre-trained multi-speaker TTS to new speakers through fine-tuning.
For the speaker adaptation approach, there are two common fine-tuning methods: fine-tune the speaker embedding only ($\{E_S\}$), or fine-tune the whole model ($\{\bm\theta_E, \bm\theta_{VA}, \bm\theta_D, E_S\}$).
However, since the speaker embedding look-up table $E_S$ only learns the embeddings of the training speakers, we need to initialize a new embedding table $\hat{E_S}$ for the testing speakers so that each testing speaker $i$ can have its corresponding speaker embedding $\hat{e_i}$.
Moreover, regarding that the encoder $\bm\theta_E$ does not and should not condition on the speaker, as mentioned in Section~\ref{ssec:multi-spk-fastspeech}, we would not update $\bm\theta_E$ during the fine-tuning stage.
In conclusion, we would either fine-tune the speaker embedding only ($\{\hat{E_S}\}$), or fine-tune the speaker embedding with both the variance adapter and the decoder ($\{\bm\theta_{VA}, \bm\theta_D, \hat{E_S}\}$).
In this paper, we mainly focus on fine-tuning $\{\bm\theta_{VA}, \bm\theta_D, \hat{E_S}\}$ in our experiments.

\section{Meta-TTS}
\label{sec:meta-tts}

\begin{algorithm}[t]
\caption{MAML for Few-Shot Learning}
\label{alg:MAML}
\begin{algorithmic}[1]
    \REQUIRE $p(\mathcal{T})$: distribution over tasks
    \REQUIRE $\alpha$, $\beta$: step size hyperparameters
    \STATE randomly initialize $\bm\theta$
    \WHILE{not done}
        \STATE Sample batch of tasks $\mathcal{T}_i \sim p(\mathcal{T})$
        \FORALL{$\mathcal{T}_i$}
            \STATE Sample $K$ datapoints $\mathcal{S}_i=\{(\mathbf{x}_j^{\mathcal{S}_i},\mathbf{y}_j^{\mathcal{S}_i})\}_{j=1}^K$ from $\mathcal{T}_i$
            \STATE Let $\bm\theta_i \gets \bm\theta$
            \FOR{$N$ gradient decent steps}
                \STATE Inner update: $\bm\theta_i \gets \bm\theta_i - \alpha \nabla_{\bm\theta_i}\mathcal{L}\left(\bm\theta_i, \mathcal{S}_i\right)$
            \ENDFOR
            \STATE Sample $Q$ datapoints $\mathcal{Q}_i=\{(\mathbf{x}_j^{\mathcal{Q}_i},\mathbf{y}_j^{\mathcal{Q}_i})\}_{j=1}^Q$ from $\mathcal{T}_i$ for the meta-update
        \ENDFOR
        \STATE Meta-update: $\bm\theta \gets \bm\theta - \beta\nabla_{\bm\theta}\mathbb{E}_{\mathcal{T}_i \sim p(\mathcal{T})}\mathcal{L}\left(\bm\theta_i, \mathcal{Q}_i\right)$
    \ENDWHILE
\end{algorithmic}
\end{algorithm}

Meta-learning, also known as ``learn to learn'', intends to design models that can learn new skills or adapt to new environments rapidly with a few training examples. Therefore, meta-learning is suitable and broadly used for solving few-shot downstream tasks.
Similar to supervised learning, which fits the model on a set of training data points and evaluates on a set of testing data points, meta-learning fits the model on a set of training tasks then evaluates on a set of testing tasks~\cite{chao2020revisiting}.
In this paper, we choose MAML~\cite{finn2017model} as our meta-learning algorithm, which strives to learn a good model meta-initialization (i.e. the initialization learned through meta-learning for further transfer learning) for the downstream tasks.
The full MAML algorithm is shown in Algorithm~\ref{alg:MAML}.


In Section~\ref{ssec:maml}, we describe the MAML training framework in a general manner. Moreover, we explain some modifications on MAML for meta-learning the multi-speaker FastSpeech 2 in Section~\ref{ssec:meta-train}.

\subsection{An introduction to MAML}
\label{ssec:maml}

Given that we are going to evaluate the model on a set of $K$-shot regression downstream tasks, MAML would generate a set of $K$-shot regression training tasks $\mathcal{T}$ from the training dataset accordingly.
In the following, we would call each task $\mathcal{T}_i$ a ``meta-task'' to indicate the task is for meta-training (i.e. meta-learning's training procedure).
As explained in Section~\ref{sec:problem-def}, each meta-task $\mathcal{T}_i$ consists of a support set $\mathcal{S}_i$ of $K$ examples and a query set $\mathcal{Q}_i$ of $Q$ examples.
The goal of each meta-task $\mathcal{T}_i$ is to learn task-specific parameters $\bm\theta_i$ using the support set $\mathcal{S}_i$ (inner loop):
\begin{equation}
    \bm\theta_i \coloneqq \argmin_{\bm\theta}\mathcal{L}
    \left(\bm\theta, \mathcal{S}_i\right).
    \label{eq:1}
\end{equation}
$\mathcal{L}(\bm\theta, \cdot)$ is the loss function of the task, where $\bm\theta$ are the model parameters before inner-loop adaptation and $\bm\theta_i$ is the task-adapted model parameters.

The goal of MAML is to learn meta-initialization of parameters $\bm\theta_{meta}^*$ that we can obtain well-performed task-adapted parameters $\bm\theta_i$ for each meta-task $\mathcal{T}_i$ (outer loop, the optimization procedure is called ``meta-update''):
\begin{equation}
    \bm{\theta}_{meta}^*\coloneqq \argmin_{\bm\theta}F(\bm\theta),
    \text{ where }
    F(\bm\theta)=\frac{1}{M}\sum_{i=1}^M
    \mathcal{L}\left(\bm\theta_i, \mathcal{Q}_i\right).
    \label{eq:2}
\end{equation}
$M$ represents the ``batch size'' of meta-tasks, and $F(\cdot)$ indicate the average of the meta-tasks' losses for each meta-update step.
In practice, inner loop optimization uses multiple gradient descent steps (as shown in Algorithm~\ref{alg:MAML}, line 7 to 9), and meta-update requires second-order gradient computation.

\subsection{Meta-training FastSpeech 2 with MAML}
\label{ssec:meta-train}

\begin{figure}
    \includegraphics[width=\linewidth]{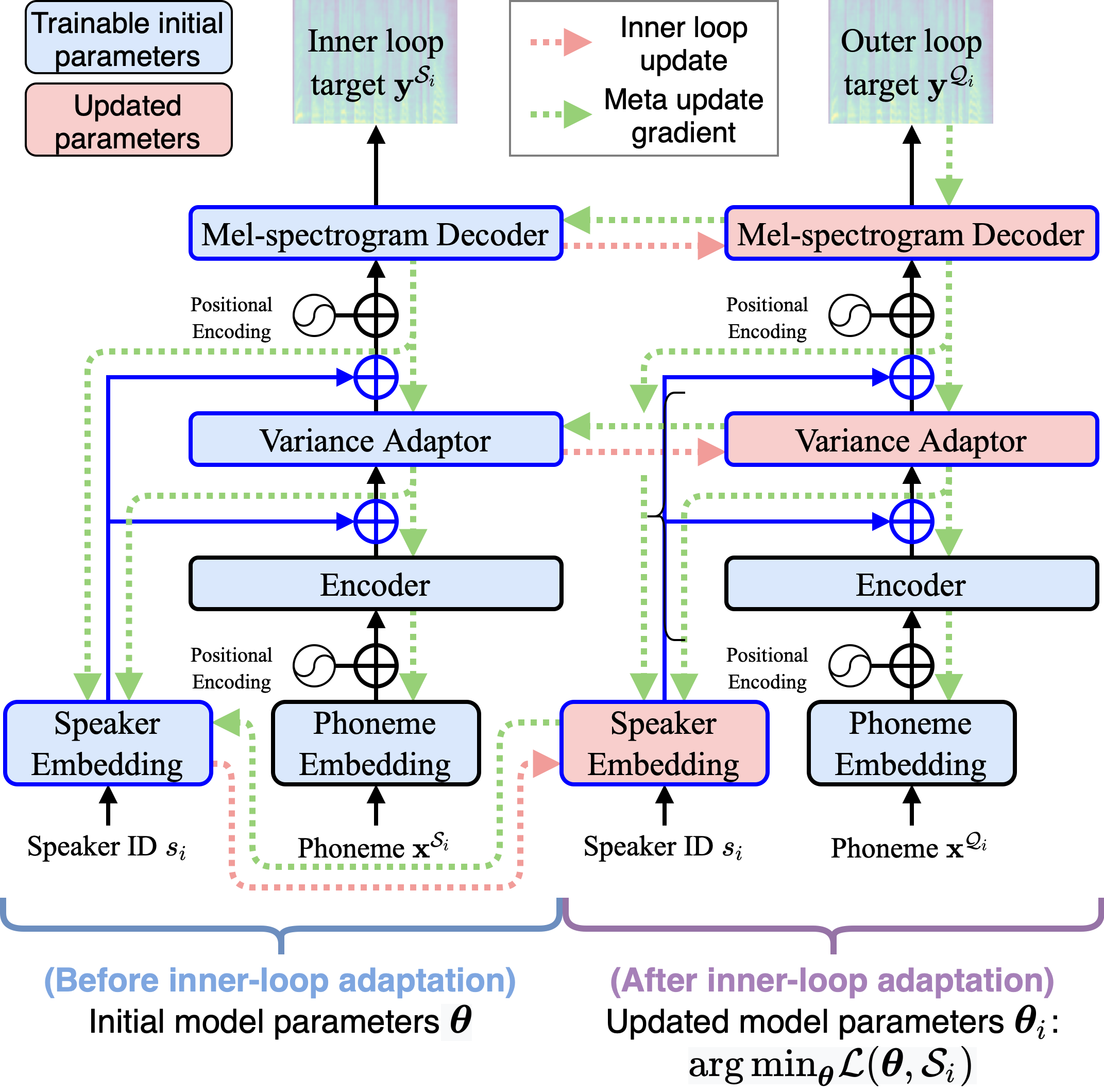}
    \caption{
    The computation graph of training MAML on a meta-task. The left half figure indicates the model parameters before an inner loop ($\bm\theta$, i.e. ``initial model parameters'' block in Figure~\ref{subfig:meta-learning}), and the right half figure indicates the adapted model parameters after an inner loop ($\bm\theta_i$, i.e. ``updated model parameters'' block in Figure~\ref{subfig:meta-learning}).
    Red dashed arrows and red blocks indicate which parameters are updated through the inner loop.
    After the inner loop adaptation, we feed-forward the input data of the query set through the adapted model (right half figure), then calculate the loss on its output (upper right figure) as the loss of the meta-task.
    Last, the loss is back-propagated (green dashed arrows) from the upper right of the figure back to the initial model parameters before the inner loop (left half figure).
    }
    \label{fig:meta-tts}
\end{figure}

Since we aim to solve $K$-shot speaker adaptation tasks\footnote{Because MAML is applied to the speaker adaptation approach in this paper, we may simply use ``speaker adaptation'' to refer to the voice cloning task/behavior in this section.}, we need to construct a set of meta-tasks from the multi-speaker training corpus, where the support and query set of each meta-task contain $K$ and $Q$ examples of the same speaker, respectively. In this paper, we set $Q=K$ during training and $Q=1$ for inference.
To sample a meta-task $\mathcal{T}_i$, we first uniformly sample a speaker from the training corpus, then $2K$ examples of the speaker are randomly selected, where $K$ examples for the support set $\mathcal{S}_i$ and $K$ for the query set $\mathcal{Q}_i$.

For training a multi-speaker FastSpeech 2 with MAML, we need to modify the MAML algorithm slightly.
The model parameters of a multi-speaker FastSpeech 2 could be represented as $\bm\theta=\{\bm\theta_E,\bm\theta_{VA},\bm\theta_D,E_S\}$, where $\bm\theta_E,\bm\theta_{VA},\bm\theta_D$, and $E_S$ indicates encoder, variance adapter, decoder and speaker embedding look-up table respectively.
As noted in Section~\ref{sec:spk-adaptation}, we would not update the encoder $\bm\theta_E$ during adaptation, and thus it should also be fixed during the inner loop adaptation.
That is, Equation~\ref{eq:1} would become:
\begin{equation}
    \bm\theta_i = \left\{\bm\theta_E, \bm\theta_{VA,i}, \bm\theta_{D,i}, E_{S,i} \right\},
    \label{eq:3}
\end{equation}
where
\begin{equation}
    \bm\theta_{VA,i}, \bm\theta_{D,i}, E_{S,i}\coloneqq \argmin_{\bm\theta_{VA}, \bm\theta_D, E_S}\mathcal{L}
    \left(\bm\theta, \mathcal{S}_i\right).
    \label{eq:4}
\end{equation}
The illustration of our meta-training process is in Figure~\ref{fig:meta-tts}.

Our modification is similar to Almost No Inner Loop (ANIL)~\cite{raghu2019rapid}, where only some modules are adapted in the inner loop, while the idea is different. The concept of ANIL is ``feature reuse''. They find out that MAML performs well because of its powerful learned representation, while only the last layer fits the new tasks. Rely on this observation, ANIL only updates the last layer in the inner loop, then meta-updates the whole model.
On the other hand, we do not observe the phenomenon of ``feature reuse'' from Meta-TTS. 
Since each module has its responsibility, we know which modules affect speaker adaptation more. So we could directly update those speaker-dependent modules in the inner loop, then meta-update the whole model in the outer loop.

\subsection{Fine-tuning}
Since MAML is just the training algorithm to find a better model initialization for fine-tuning, the fine-tuning procedure of Meta-TTS is the same as multi-speaker TTS's in Section~\ref{sec:spk-adaptation}.
Notice that since the inner loop's goal is to mimic the fine-tuning process, the updated modules should be the same as the ones for fine-tuning.
For instance, if we only fine-tune the speaker embedding table during inference, we should only fine-tune the speaker embedding table in the inner loops.

\section{Other variants}
\label{sec:variants}

\subsection{Sharing embedding across all speakers before adaptation}
\label{ssec:share-emb}

For both our baseline model (i.e. multi-speaker TTS mentioned in Section~\ref{sec:multi-spk-tts} and~\ref{sec:spk-adaptation}) and Meta-TTS, there is a common problem for adapting to an unseen speaker: we need to re-initialize the speaker embedding for the new speaker.
It is somehow inevitable for transfer learning from the baseline model, since it is reasonable for a new task to tune on new parameters.
However, it is not that reasonable for meta-learning. As mentioned in Section~\ref{sec:meta-tts}, meta-learning learns from a set of training tasks, then wish the learned capability of ``adapting on a task'' can be utilized on new downstream tasks, i.e. testing tasks.
But here comes to the problem: the model learns the capability of adapting on a task with its speaker embedding meta-trained, not with it randomly initialized.
This might cause significant degradation to the model's capability of adaptation.

To deal with the problem, we propose an architecture variant: for the models trained with both algorithms (multi-task/meta training), we do not use the speaker embedding look-up table $E_S$ anymore. Instead, we construct a shared speaker embedding $e_S$.
That is, after multi-task/meta training, the trained shared speaker embedding $e_S$ would become a speaker embedding (meta-)initialization, which can be further fine-tuned on any speaker adaptation task and become a better speaker-dependent embedding, including new tasks with unseen speakers.
Although the shared speaker embedding does not suffer from the re-initialization problem, it loses the speaker embedding capacity, which is a trade-off.
Overall, both variants (using speaker embedding look-up table or using shared speaker embedding) have their pros and cons, and we will evaluate their performance and analyze their difference in our experiments.

\subsection{Fine-tuning different modules}
\label{ssec:different-modules}
While we need to update the speaker embedding $\hat{E_S}$ and not the encoder $\bm\theta_E$ for fine-tuning, it is unclear that whether updating the variance adaptor and the decoder would lead to better results.
Therefore, we also experiment with fine-tuning different module combinations.
Besides fine-tuning $\{\hat{E_S}\}$ and $\{\bm\theta_{VA}, \bm\theta_D, \hat{E_S}\}$, we also try to tune $\{\bm\theta_D, \hat{E_S}\}$ and $\{\bm\theta_{VA}, \hat{E_S}\}$.
For Meta-TTS, since the inner update process should match the fine-tuning stage, we need to train a new model for each combination.

\section{Evaluation metrics}
\label{sec:evaluation-metrics}
The Mean Opinion Score (MOS) test is a well-known subjective metric to estimate synthesized speech quality. However, it is expensive and time-consuming due to the entanglement of human judges. On the other hand, although neural evaluations can not fully replace human ratings, they can be easily applied. For example, we can utilize $d$-vectors to evaluate the voice cloning performance. Also, we can use trained MOS prediction networks to estimate the audio quality.
Therefore, we only evaluate the human ratings on a randomly sampled subset of the testing set in this paper. We further assess the complete testing set with neural evaluation metrics as reference.

In Section~\ref{ssec:human-evaluation}, we will briefly describe the human evaluation metrics. Then we will explain what $d$-vector is in Section~\ref{ssec:d-vector} and the neural evaluation metrics we use in Section~\ref{ssec:neural-metrics}.

\subsection{Human evaluation (MOS, SMOS)}
\label{ssec:human-evaluation}
\subsubsection{Speaker similarity (SMOS)}
We provide a similarity MOS (SMOS) test as one of the speaker similarity metrics.
We pair each testing speech (real/synthesized) with a real utterance of the same speaker, and at least 5 judges would rate the similarity on a five-point Likert Scale (1: Bad, 2: Poor, 3: Fair, 4: Good, 5: Excellent).
We first average the ratings of each utterance then calculate each model's mean score and confidence interval.

\subsubsection{Naturalness (MOS)}
Besides speaker similarity, we also measure the audio quality of the utterances with the Mean Opinion Score (MOS).
Human raters judge each testing utterance according to its naturalness.
Unlike the SMOS settings, we do not need to pair each testing speech with an enrollment utterance.

\subsection{$D$-vector}
\label{ssec:d-vector}
$D$-vector~\cite{variani2014deep,heigold2016end} is the utterance-level feature extracted from a pre-trained speaker verification model.
The training objective of the model, GE2E~\cite{wan2018generalized}, is to increase the cosine similarity of the $d$-vectors between the utterances of the same speaker and decrease the similarity between different speakers.
Thus, the cosine similarity of two $d$-vectors indicates the speaker similarity of two utterances.
Moreover, the average of the $d$-vectors of a speaker's utterances could become a speaker representation, and the cosine similarity of two speaker representations thus denotes the similarity of two speakers.

We use a GE2E model~\cite{wan2018generalized} pre-trained by~\cite{jia2018transfer}, which consists of 3 LSTM layers followed by a projection layer. The model was trained on a combination of subsets of the three datasets: Librispeech \texttt{train-other-500}~\cite{panayotov2015librispeech}, VoxCeleb1 \texttt{Dev A-D}~\cite{nagrani2017voxceleb}, and VoxCeleb2 \texttt{Dev A-H}~\cite{chung2018voxceleb2}, which is 3201 hours of utterances in total from 8371 speakers\footnote{More pre-training details: \url{https://blue-fish.github.io/experiments/RTVC-7.html}.}.

\subsection{Neural evaluation metrics}
\label{ssec:neural-metrics}

\subsubsection{Speaker similarity}
\label{sssec:speaker-similarity}
One of the neural evaluation metrics is the speaker similarity between our synthesized speech and the real utterances of the same target speaker.
Following the standard practice in text-independent speaker verification, we select real utterances from all target speakers as the enrollment set and compute the centroid of each speaker's $d$-vector utterance embeddings as its representation.
Then for each synthesized speech, we calculate the cosine similarity between its $d$-vector and its target speaker's representation to indicate whether it is similar to the target speaker.
To evaluate the performance of an adapted TTS model, we can average the similarities of its synthesized utterances to the target speaker, where higher similarity indicates better voice cloning results.

\subsubsection{Speaker verification}
To certify whether the generated utterances belong to the correct speaker, we evaluate the ability of a speaker verification system to distinguish the generated utterances from other speakers.
For each synthesized utterance of speaker $i$, we compute the cosine similarity of its $d$-vector embedding to a randomly selected real enrollment utterance's $d$-vector of speaker $j$.
When the similarity exceeds a threshold, we will recognize the utterance as one from the same speaker ($i=j$), and vice versa.
Then we can plot the results into a detection error trade-off (DET) graph and use the equal error rate (EER) as the quantitative evaluation metric.

\subsubsection{Synthesized speech detection}
We also evaluate whether the synthesized utterances are hard to discern from the real ones by a neural model.
Similar to the speaker verification process, we would pair each synthesized/real utterance with a real enrollment utterance and compute their $d$-vectors' cosine similarity.
Nevertheless, we only sample the real enrollment utterances from the same speaker in the synthesized versus real detection task.
When the cosine similarity exceeds a threshold, we would recognize the utterance as real; otherwise, recognize it as synthesized speech.
By modifying the threshold, we can plot a receiver operating characteristic (ROC) curve according to the resulting true-positive rates and the false-positive rates. Moreover, we can use the area under the ROC curve (ROC AUC) as the quantitative evaluation metric.

\subsubsection{Naturalness MOS prediction}
Since human evaluation requires vast labor costs, there are lots of works investigating how to predict MOS through neural networks~\cite{lo2019mosnet,leng2021mbnet,tseng2021utilizing}.
The training target of MOSNet~\cite{lo2019mosnet} is to predict the average MOS of each utterance, while MBNet~\cite{leng2021mbnet} also considers the judging bias of each rater.
On the other hand, \cite{tseng2021utilizing} utilizes several typical self-supervised models for downstream MOS prediction, including Wav2vec 2.0~\cite{baevski2020wav2vec}, TERA~\cite{liu2021tera} and CPC~\cite{oord2018representation}.
In the experiments, we use the five MOS prediction networks (MOSNet, MBNet, \cite{tseng2021utilizing} with three different self-supervised models) as the judges to estimate the MOS ratings.

\section{Experimental setup of speaker adaptation}
\label{sec:exp-setup}
As mentioned in the second paragraph of Section~\ref{sec:intro}, there are two general approaches to deal with the voice cloning task: speaker adaptation and speaker encoding.
While in Section~\ref{sec:meta-tts} we apply MAML to the speaker adaptation approach, we still compare our proposed method with baselines of both approaches.
In this section, we will describe the experimental setup of comparing our proposed method with the speaker adaptation baseline, and the results are in Section~\ref{sec:exp}.
Then we will show the experimental setup and results of comparing our proposed method with the speaker encoding baseline in Section~\ref{sec:speaker-encoding-exp}.

Our FastSpeech 2 implementation is based on an unofficial GitHub repository\footnote{\url{https://github.com/ming024/FastSpeech2}}, which mostly follows the original paper but with minor modifications that stabilize the training.
In the experiments, we mainly compare the results of 5-shot speaker adaptation tasks\footnote{In Section~\ref{sec:exp-setup} and~\ref{sec:exp} we use ``speaker adaptation tasks'' to refer to voice cloning tasks.} on two models: multi-task learned multi-speaker TTS (baseline model\footnote{We use ``baseline model'' to indicate the speaker adaptation baseline throughout the whole paper, while other names are used for the speaker encoding baseline.}) from Section~\ref{sec:multi-spk-tts} and meta-learned multi-speaker TTS (Meta-TTS) from Section~\ref{sec:meta-tts}.
For Meta-TTS, we have two kinds of speaker embeddings (embedding table $E_S$ and shared embedding $e_S$ from Section~\ref{ssec:share-emb}) and four different fine-tuning module combinations (from Section~\ref{ssec:different-modules}).
Although it is not reasonable for the baseline model to use shared embedding, we also apply the two speaker embedding types and four fine-tuning module combinations to the baseline model for a fair comparison.
To note, the trained baseline model can directly apply different fine-tuning module combinations, while Meta-TTS needs to train separately for each combination. 

\paragraph{Datasets}
We train our models on \texttt{train-clean-100} subset of LibriTTS~\cite{zen2019libritts} dataset, which contains around 54 hours of utterances from 247 speakers.
To evaluate the custom voice scenario, we not only test our pre-trained models on LibriTTS \texttt{test-clean} subset but also adapt them to voices in another corpus VCTK~\cite{yamagishi2019cstr}.
LibriTTS \texttt{test-clean} subset contains 8.56 hours from 39 speakers with an average of 13 minutes per speaker, and VCTK contains 44 hours from 110 speakers with around 400 clean speech recordings each.

\paragraph{Training}
Since we are evaluating 5-shot speaker adaptation tasks, each meta-task would only contain 10 utterances from a speaker during training, 5 for the support set, and 5 for the query set.
In each inner loop during meta-training, we update the model with 5 iterations. We then averaged the gradients of 8 meta-tasks ($M$ in Equation~\ref{eq:2}) for each meta-update step since we train the Meta-TTS with multi-GPU training distributing across meta-tasks on an 8-GPU device.
Also, we had experimented with setting $M$ to 1 and 4. While the training loss of Meta-TTS can decrease gradually, their resulting audio quality is worse than setting $M$ to 8, so we only show the results of setting $M=8$.
Therefore there are 80 training samples used for a meta-update step in total, so to be fair in comparison, the batch size of the baseline model is also set to 80.
Moreover, since the inner loop updates are only to calculate the loss of the outer loop ($F$ in Equation~\ref{eq:2}), we should set the number of meta-steps of Meta-TTS equal to the training steps of the baseline model for a reasonable comparison.
Therefore, we train both models for 100k (meta-)steps.

\paragraph{Inference}
We uniformly sample 16 tasks for each speaker, where each task contains 5 utterances for the support set and 1 for the query set.
For each task, we fine-tune the (meta-)trained models with the support set, then generate the synthesized utterance from the inputs of the query set.
The targets of the query sets are the upper bound references.
Besides, since human evaluations require colossal labor costs, we further sample a subset from the test set for human evaluations: 30 speakers from LibriTTS \texttt{test-clean} set and 80 speakers from VCTK, with 1 random task per speaker.
To determine how many steps the models need for adaptation, we adapt them with different fine-tuning steps to get their adaptation trends.
The adapted models' output mel-spectrograms will be converted to waveforms by a vocoder before evaluation.
To ensure that the target speakers' characteristics of the generated speeches are all coming from the TTS model and not from the vocoder, we choose a pre-trained vocoder which is not trained on our testing data before, especially the testing speakers should be unseen.
Therefore, we use MelGAN~\cite{kumar2019melgan} as our neural vocoder during inference, which is pre-trained on an internal 6-speaker dataset with 3 male and 3 female speakers by the MelGAN's authors.
Since our testing speakers are unseen during the vocoder training, generalization problems might exist, which might affect the speech quality, and also the vocoder might not preserve the target speaker's characteristics of the mel-spectrogram well.
Thus we also use the vocoder to convert ground truth mel-spectrograms as a reference baseline (noted as the ``reconstructed'' utterances), which should be the upper bound of our synthesized speech.
Moreover, the gap between the ``reconstructed'' and the ground truth utterances indicates the vocoder's generalization gap.

\section{Speaker adaptation results}
\label{sec:exp}

\subsection{Speaker similarity}
\label{ssec:similarity}

\begin{table*}[!h]
  \caption{Speaker similarity evaluated by similarity MOS (SMOS) with 95\% confidence intervals.}
  \label{tab:smos}
  \centering
  \begin{tabular}{llcccc}
    \toprule
    \multirow{2}{*}{Approach}  & \multirow{2}{*}{Adaptation} & \multicolumn{2}{c}{LibriTTS} & \multicolumn{2}{c}{VCTK}\\
    \cmidrule(lr){3-4} \cmidrule(lr){5-6}
    & & Emb table $\hat{E_S}$ & Shared $e_S$ & Emb table $\hat{E_S}$ & Shared $e_S$\\
    \midrule
    \multicolumn{2}{l}{Real} 
        & \multicolumn{2}{c}{$4.29 \pm 0.27$} & \multicolumn{2}{c}{$4.54 \pm 0.09$} \\
    \multicolumn{2}{l}{Reconstructed} 
        & \multicolumn{2}{c}{$3.33 \pm 0.29$} & \multicolumn{2}{c}{$4.08 \pm 0.12$} \\
    \midrule
    Baseline & 10 steps
        & $1.53 \pm 0.18$ & $1.34 \pm 0.21$
        & $1.56 \pm 0.12$ & $1.32 \pm 0.13$\\
    Meta-TTS & 10 steps
        & $\mathbf{2.77 \pm 0.24}$ & $\mathbf{2.67 \pm 0.28}$
        & $\mathbf{3.14 \pm 0.16}$ & $\mathbf{3.45 \pm 0.14}$\\
    \bottomrule
  \end{tabular}
\end{table*}

This section aims to measure the speaker similarity of the speech synthesized from the adapted models.
Since human evaluation requires enormous labor costs, we only evaluate the results of fine-tuning the whole model (except for the encoder) with 10 adaptation steps to show the effectiveness of our proposed Meta-TTS.
%
The similarity Mean Opinion Score (SMOS) results are in Table~\ref{tab:smos}.
We also evaluate the SMOS of the (reconstructed) real utterances as the similarity upper bound.
As shown, Meta-TTS beats the baseline model with more than 1 point under all the settings.
Since it is not reasonable for the baseline model to use shared speaker embedding, it is expectable that the SMOS would become worse.
However, Meta-TTS with shared speaker embedding obtains comparable results on LibriTTS and better on VCTK.
From these results, we can probably infer that for Meta-TTS with shared speaker embedding $e_S$, because all of the fine-tuning modules ($\{\bm\theta_{VA},\bm\theta_D,e_S\}$) are meta-trained, they are suitable to fine-tune quickly.
On the other hand, when using speaker embedding table $\hat{E_S}$, since the testing speakers' embedding table is not meta-trained, it might somehow reduce the adaptation effectiveness.

\begin{figure*}[!h]
    \centering
    \includegraphics[width=\linewidth]{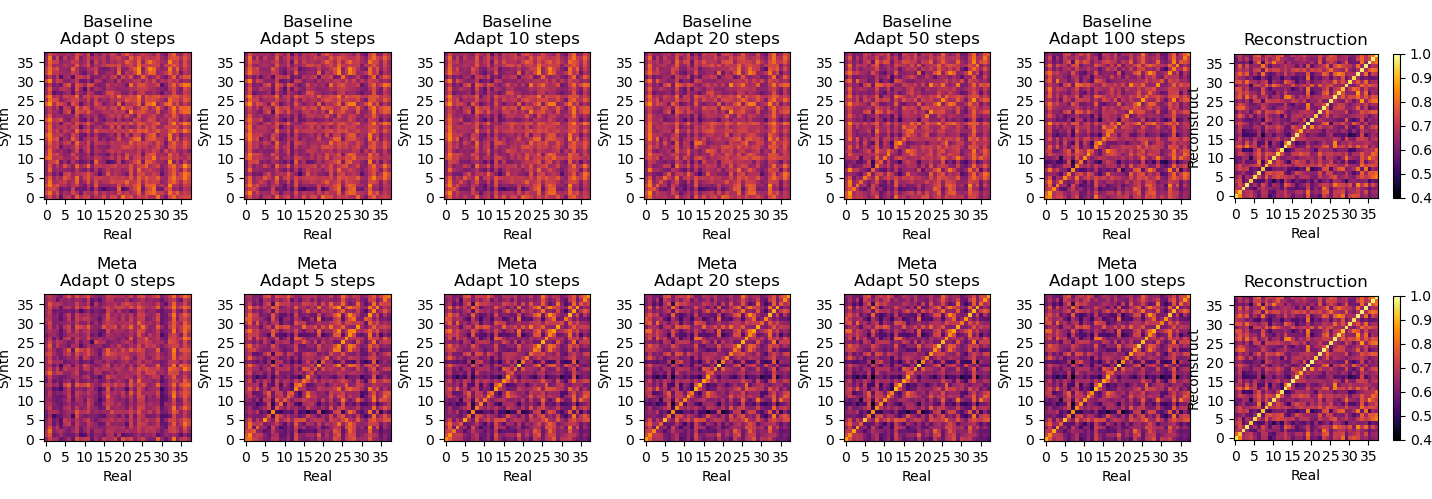}
    \caption{Cosine similarity matrices calculated on LibriTTS \texttt{test-clean}. The first row are the results of the baseline model, and the second row are the results of Meta-TTS. Each column represents different adaptation steps, while the last column is the result of the reconstructed utterances (two matrices are the same), which is plotted as the final adaptation goal.}
    \label{fig:spk_sim}
\end{figure*}

\begin{figure*}[!h]
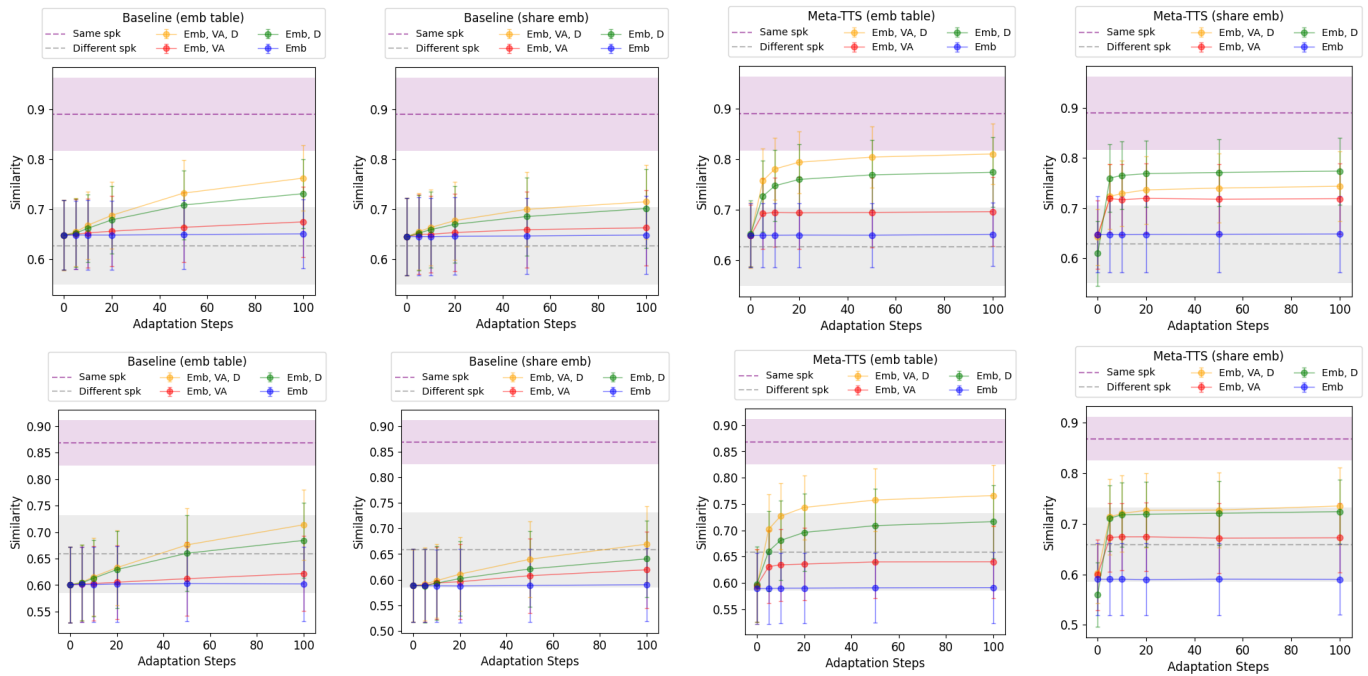

    \graphicspath{{LibriTTS/}}
    \DeclareGraphicsExtensions{.pdf,.jpeg,.png}
    \centering
    \begin{subfigure}[]{0.24\linewidth}
        \includegraphics[width=\linewidth]{errorbar_plot_base_emb}
    \end{subfigure}
    \hfill
    \begin{subfigure}[]{0.24\linewidth}
        \includegraphics[width=\linewidth]{errorbar_plot_base_emb1}
    \end{subfigure}
    \hfill
    \centering
    \begin{subfigure}[]{0.243\linewidth}
        \includegraphics[width=\linewidth]{errorbar_plot_meta_emb}
    \end{subfigure}
    \hfill
    \begin{subfigure}[]{0.243\linewidth}
        \includegraphics[width=\linewidth]{errorbar_plot_meta_emb1}
    \end{subfigure}
    \\

    \graphicspath{{VCTK/}}
    \centering
    \begin{subfigure}[]{0.24\linewidth}
        \includegraphics[width=\linewidth]{errorbar_plot_base_emb}
    \end{subfigure}
    \hfill
    \begin{subfigure}[]{0.24\linewidth}
        \includegraphics[width=\linewidth]{errorbar_plot_base_emb1}
    \end{subfigure}
    \hfill
    \centering
    \begin{subfigure}[]{0.243\linewidth}
        \includegraphics[width=\linewidth]{errorbar_plot_meta_emb}
    \end{subfigure}
    \hfill
    \begin{subfigure}[]{0.243\linewidth}
        \includegraphics[width=\linewidth]{errorbar_plot_meta_emb1}
    \end{subfigure}
    \caption{Speaker similarity results of the two models (baseline/Meta-TTS) with two types of speaker embedding and four fine-tuning module combinations on the two corpora (LibriTTS/VCTK). The colored regions / error bars represent the standard deviations. \textbf{Top row:} LibriTTS. \textbf{Bottom row:} VCTK. \textbf{Emb}: speaker embedding. \textbf{VA}: variance adaptor. \textbf{D}: decoder.}
    \label{fig:spk-similarity}
\end{figure*}

As stated in Section~\ref{sssec:speaker-similarity}, we can further evaluate the speaker adaptation results with the $d$-vector cosine similarities.
For each target speaker, we can average the synthesized utterances' $d$-vectors as its synthesized speaker representation.
On the other hand, we can average each speaker's real utterances' $d$-vectors as its real speaker representation.
Then we show the cross-speaker similarity between synthesized representations and the real representations in Figure~\ref{fig:spk_sim}, where the numbers in both axes are the speaker identities.
The upper row is the results of the baseline model with different adaptation steps, and the lower row is the results of Meta-TTS, where both models use the speaker embedding table and fine-tune the whole model (except the encoder).
The rightmost column of both rows is the same, which is the ``reconstructed'' versus ``ground truth'' similarity result and serves as the upper bound of speaker adaptation.
From the rightmost column, we can observe that an excellent speaker-adapted model should have higher diagonal values and lower values otherwise, which means a well-adapted model should generate sounds close to the target speaker and far from the others.
While this pattern gradually appears after 50 adaptation steps for the baseline model, we can find this pattern with 5 to 10 adaptation steps only for Meta-TTS.
This indicates that Meta-TTS generates voices similar to the target speakers after 5 adaptation steps, while the baseline model starts after 50 adaptation steps.

To give a more statistical evaluation result, we also compute the cosine similarity between each synthesized utterance and its target speaker representation, then calculate their mean and variance.
The results are in Figure~\ref{fig:spk-similarity}.
First, we can see that Meta-TTS can adapt quickly by fine-tuning the whole model (speaker embedding + variance adaptor + decoder), which beats the baseline model of all settings.
Next, let us focus on the Meta-TTS results.
The similarity curves show that Meta-TTS with speaker embedding table does not always perform better or worse than with shared embedding table.
The main reason is that the speaker embedding table method requires re-initializing a new embedding table for the testing speakers, thus affecting the fine-tuning results.
Furthermore, we can observe that Meta-TTS with shared speaker embedding saturates quickly, while Meta-TTS with speaker embedding table can keep improving.
We will explain the reason for this phenomenon later.

\subsection{Speaker verification}

\begin{figure*}[!h]
    \centering
    \begin{subfigure}[]{0.24\linewidth}
        \includegraphics[width=\linewidth]{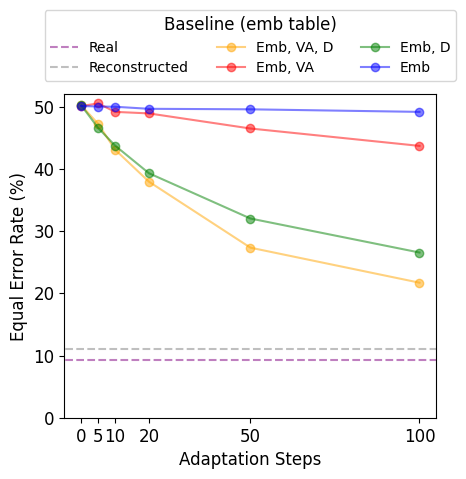}
    \end{subfigure}
    \hfill
    \begin{subfigure}[]{0.24\linewidth}
        \includegraphics[width=\linewidth]{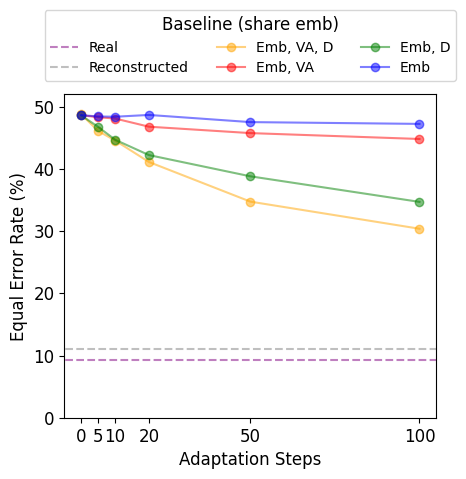}
    \end{subfigure}
    \hfill
    \begin{subfigure}[]{0.24\linewidth}
        \includegraphics[width=\linewidth]{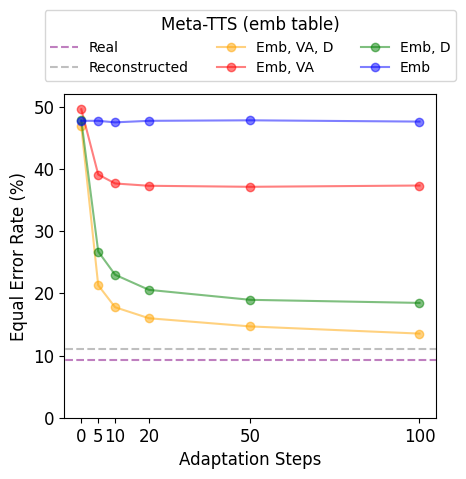}
    \end{subfigure}
    \hfill
    \begin{subfigure}[]{0.24\linewidth}
        \includegraphics[width=\linewidth]{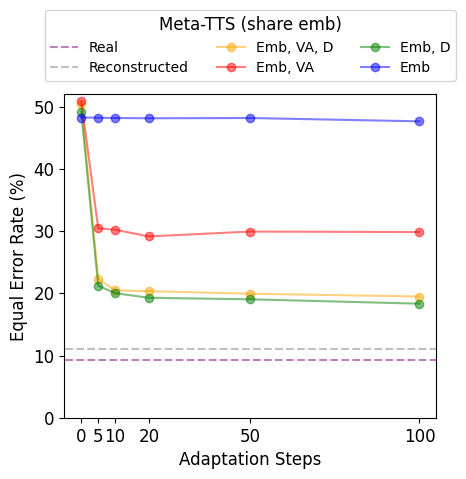}
    \end{subfigure}
    \\
    \centering
    \begin{subfigure}[]{0.24\linewidth}
        \includegraphics[width=\linewidth]{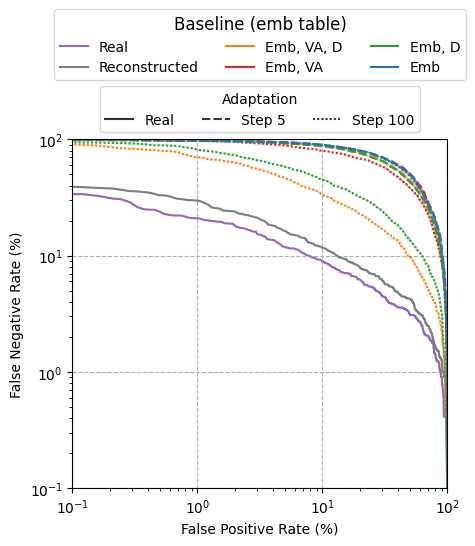}
    \end{subfigure}
    \hfill
    \begin{subfigure}[]{0.24\linewidth}
        \includegraphics[width=\linewidth]{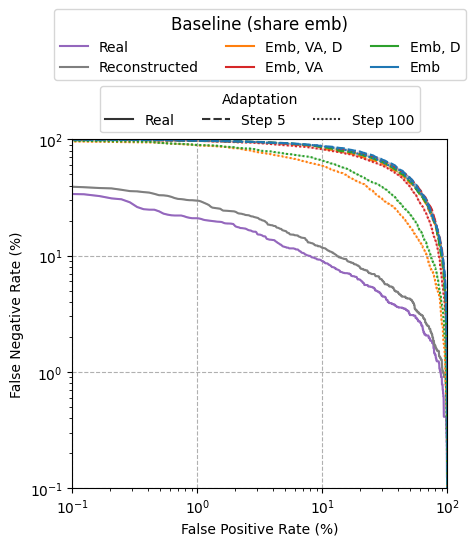}
    \end{subfigure}
    \hfill
    \begin{subfigure}[]{0.24\linewidth}
        \includegraphics[width=\linewidth]{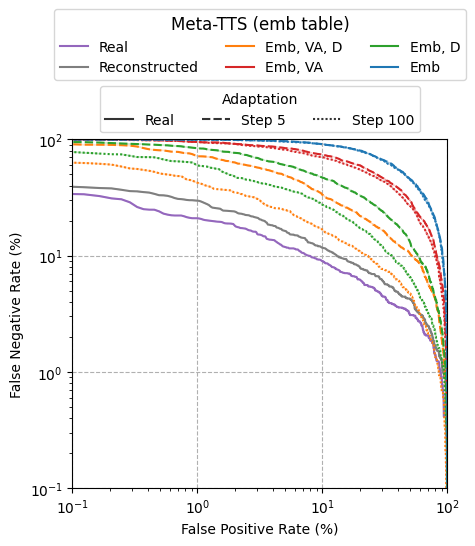}
    \end{subfigure}
    \hfill
    \begin{subfigure}[]{0.24\linewidth}
        \includegraphics[width=\linewidth]{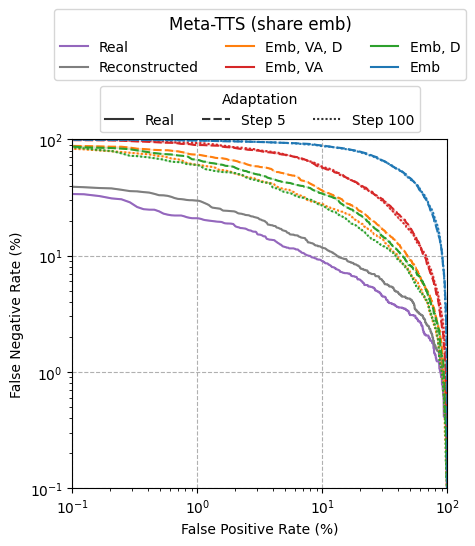}
    \end{subfigure}
    \caption{Speaker verification results on LibriTTS. \textbf{Top row:} EER. \textbf{Bottom row:} DET curves.}
    \label{fig:sv-LibriTTS}
\end{figure*}

\begin{figure*}[!h]
    \centering
    \begin{subfigure}[]{0.24\linewidth}
        \includegraphics[width=\linewidth]{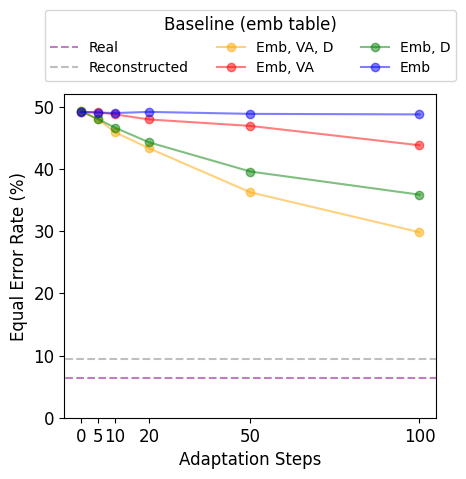}
    \end{subfigure}
    \hfill
    \begin{subfigure}[]{0.24\linewidth}
        \includegraphics[width=\linewidth]{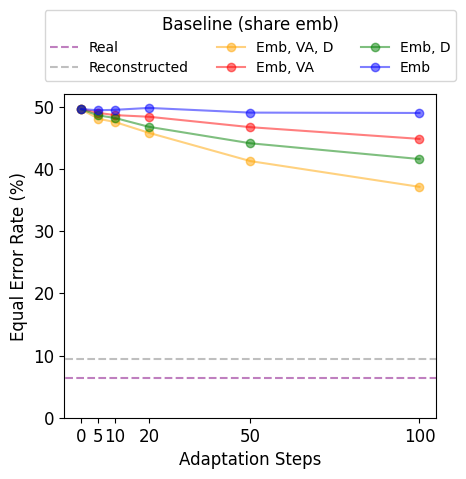}
    \end{subfigure}
    \hfill
    \centering
    \begin{subfigure}[]{0.24\linewidth}
        \includegraphics[width=\linewidth]{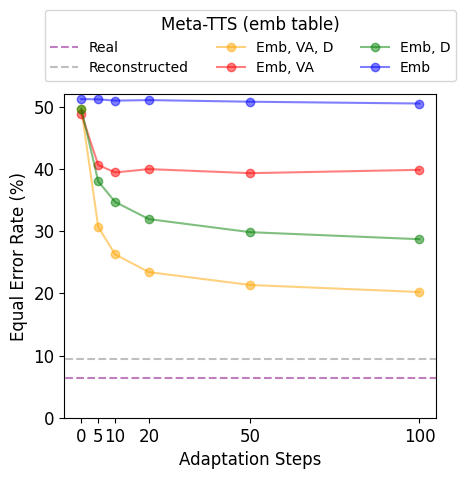}
    \end{subfigure}
    \hfill
    \begin{subfigure}[]{0.24\linewidth}
        \includegraphics[width=\linewidth]{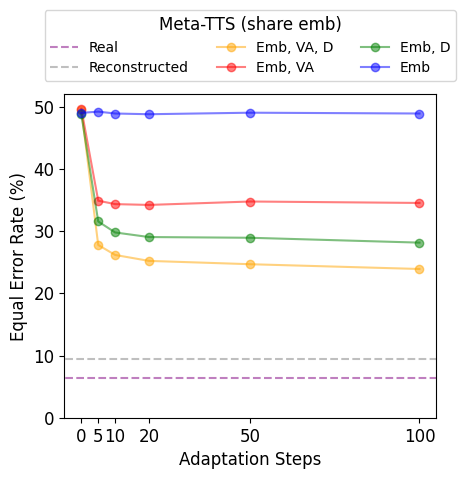}
    \end{subfigure}
    \\
    \centering
    \begin{subfigure}[]{0.24\linewidth}
        \includegraphics[width=\linewidth]{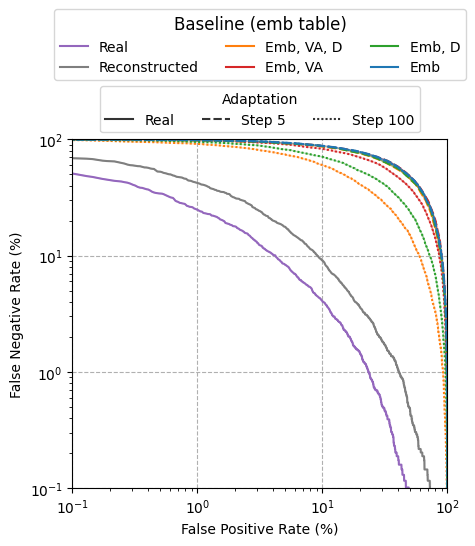}
    \end{subfigure}
    \hfill
    \begin{subfigure}[]{0.24\linewidth}
        \includegraphics[width=\linewidth]{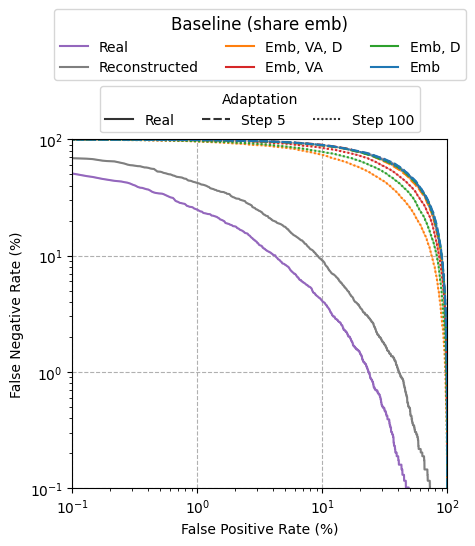}
    \end{subfigure}
    \hfill
    \centering
    \begin{subfigure}[]{0.24\linewidth}
        \includegraphics[width=\linewidth]{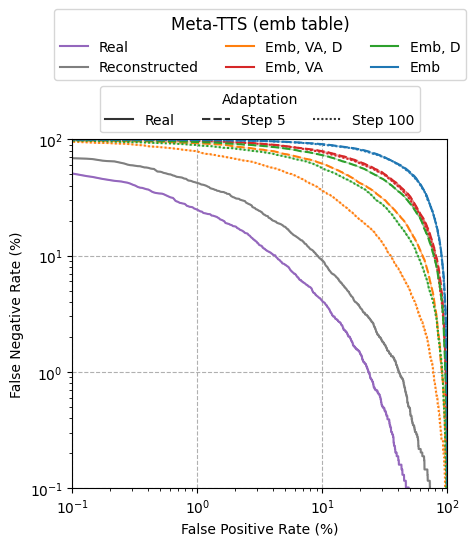}
    \end{subfigure}
    \hfill
    \begin{subfigure}[]{0.24\linewidth}
        \includegraphics[width=\linewidth]{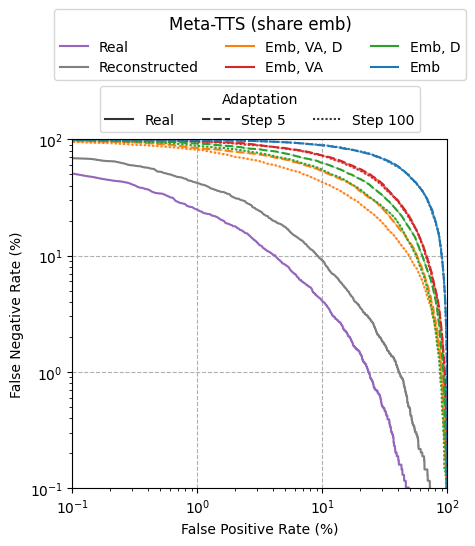}
    \end{subfigure}
    \caption{Speaker verification results on VCTK. \textbf{Top row:} EER. \textbf{Bottom row:} DET curves.}
    \label{fig:sv-VCTK}
\end{figure*}

While similar to the speaker similarity, speaker verification requires the synthesized utterances to be distinguishable from other speakers.
Equal error rate (EER, lower is better) and detection error trade-off (DET, bottom left is better) curve for the speaker verification results are shown in Figure~\ref{fig:sv-LibriTTS} and~\ref{fig:sv-VCTK}.
We also plot the (reconstructed) real utterances' results as the upper bound, where we could also observe the gap between grey and purple curves as the vocoder's generalization gap.
Overall, the observations are the same as the speaker similarity evaluation: Meta-TTS with fine-tuning the whole model (speaker embedding + variance adaptor + decoder) performs the best, and Meta-TTS with shared speaker embedding saturates quickly.
To dig into this intriguing phenomenon, we visualize their $d$-vector utterance embeddings in the following subsection.

\subsection{Visualizing $d$-vector utterance embeddings}
\label{ssec:visualize}

\begin{figure}[!h]
    \centering
    \begin{subfigure}[]{0.49\linewidth}
        \includegraphics[width=\linewidth]{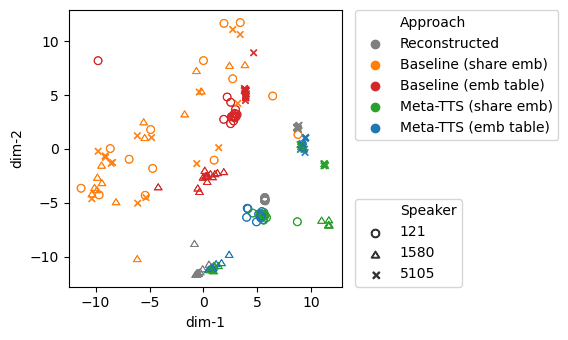}
        \caption{LibriTTS}
    \end{subfigure}
    \begin{subfigure}[]{0.49\linewidth}
        \includegraphics[width=\linewidth]{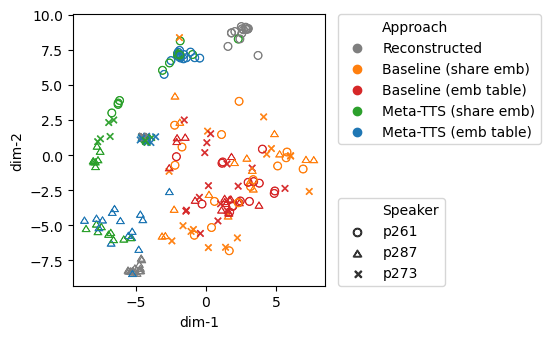}
        \caption{VCTK}
    \end{subfigure}
    \caption{t-SNE visualization of the $d$-vector utterance embeddings.}
    \label{fig:tsne}
\end{figure}

To give a direct insight into the adaptation results, we visualize the $d$-vector utterance embeddings of the generated speech in Figure~\ref{fig:tsne}.
All models adapt 20 steps by fine-tuning the whole model (speaker embedding + variance adaptor + decoder).
As shown, the baseline models' utterance embeddings are spread all over the embedding space.
In contrast, the embeddings of Meta-TTS are generally much closer to their targets (embeddings of the reconstructed real utterances), which implies that Meta-TTS adapts much more efficiently than the baseline models.
Next, please focus on the embeddings of Meta-TTS. We could find out that the embeddings of Meta-TTS with $\hat{E_S}$ are relatively compact, while those of Meta-TTS with $e_S$ have some outliers scattered elsewhere.
The following is our speculation.
When using embedding table $E_S$, the job of each module is well defined: speaker embedding table $E_S$ for saving general speaker information; variance adapter $\bm\theta_{VA}$ for predicting the duration, pitch, and energy; and decoder $\bm\theta_D$ for deciding the timbre.
However, when using shared embedding $e_S$, since it could not save any individual speaker information, the variance adapter and the decoder would be more sensitive and vulnerable to the input data to fit the training tasks.
Thus, since few-shot tasks are usually varying, the adaptation results would be more unstable for fine-tuning.
Therefore, there would be more outliers and further affect the result of speaker verification.
We can also apply the conclusion to the baseline models, whose utterance embeddings spread more widely with shared embedding $e_S$ than with embedding table $\hat{E_S}$.

\subsection{Synthesized speech detection}

\begin{figure*}[!h]
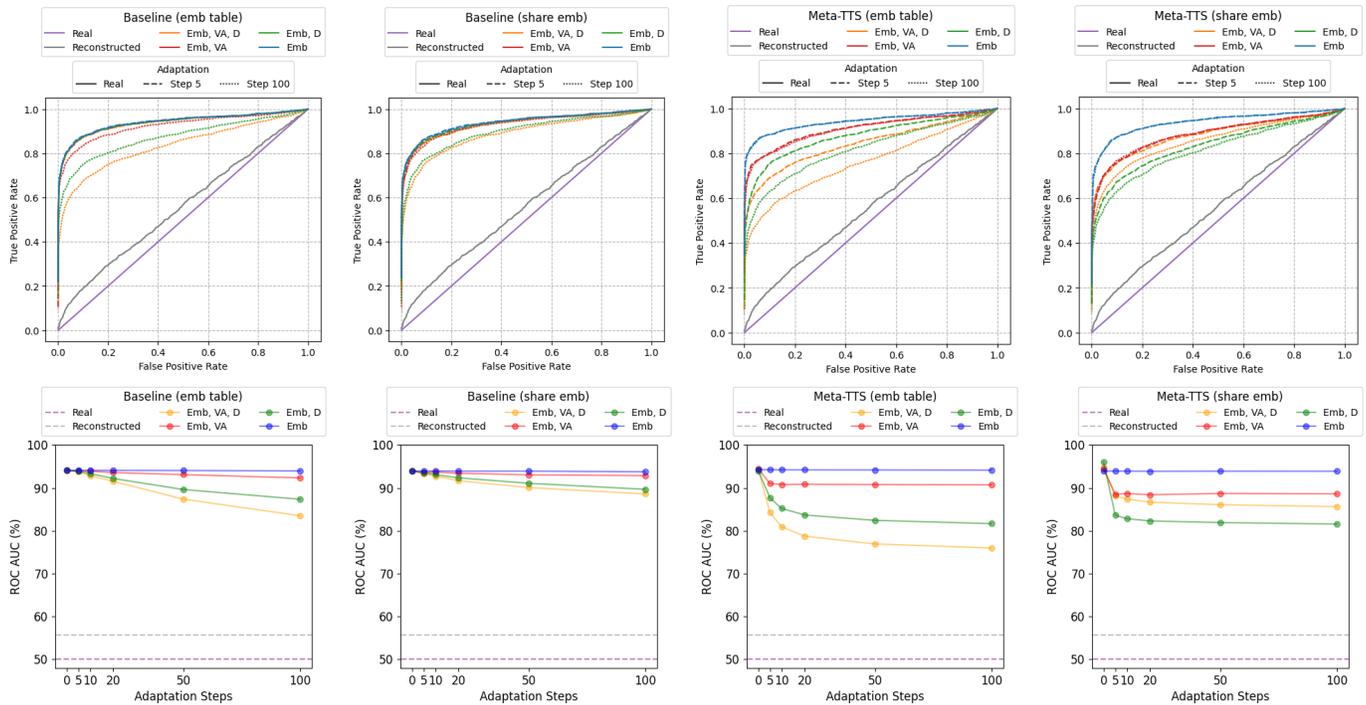

    \graphicspath{{LibriTTS/}}
    \centering
    \begin{subfigure}[]{0.24\linewidth}
        \includegraphics[width=\linewidth]{roc_base_emb.png}
    \end{subfigure}
    \hfill
    \begin{subfigure}[]{0.24\linewidth}
        \includegraphics[width=\linewidth]{roc_base_emb1.png}
    \end{subfigure}
    \hfill
    \centering
    \begin{subfigure}[]{0.243\linewidth}
        \includegraphics[width=\linewidth]{roc_meta_emb.png}
    \end{subfigure}
    \hfill
    \begin{subfigure}[]{0.243\linewidth}
        \includegraphics[width=\linewidth]{roc_meta_emb1.png}
    \end{subfigure}
    \\
    \centering
    \begin{subfigure}[]{0.24\linewidth}
        \includegraphics[width=\linewidth]{auc_base_emb.png}
    \end{subfigure}
    \hfill
    \begin{subfigure}[]{0.24\linewidth}
        \includegraphics[width=\linewidth]{auc_base_emb1.png}
    \end{subfigure}
    \hfill
    \centering
    \begin{subfigure}[]{0.24\linewidth}
        \includegraphics[width=\linewidth]{auc_meta_emb.png}
    \end{subfigure}
    \hfill
    \begin{subfigure}[]{0.24\linewidth}
        \includegraphics[width=\linewidth]{auc_meta_emb1.png}
    \end{subfigure}
    \caption{Synthesized speech detection results on LibriTTS. \textbf{Top row:} ROC curves. \textbf{Bottom row:} ROC AUC.}
    \label{fig:synthesize-detection-LibriTTS}
\end{figure*}
\begin{figure*}[!h]
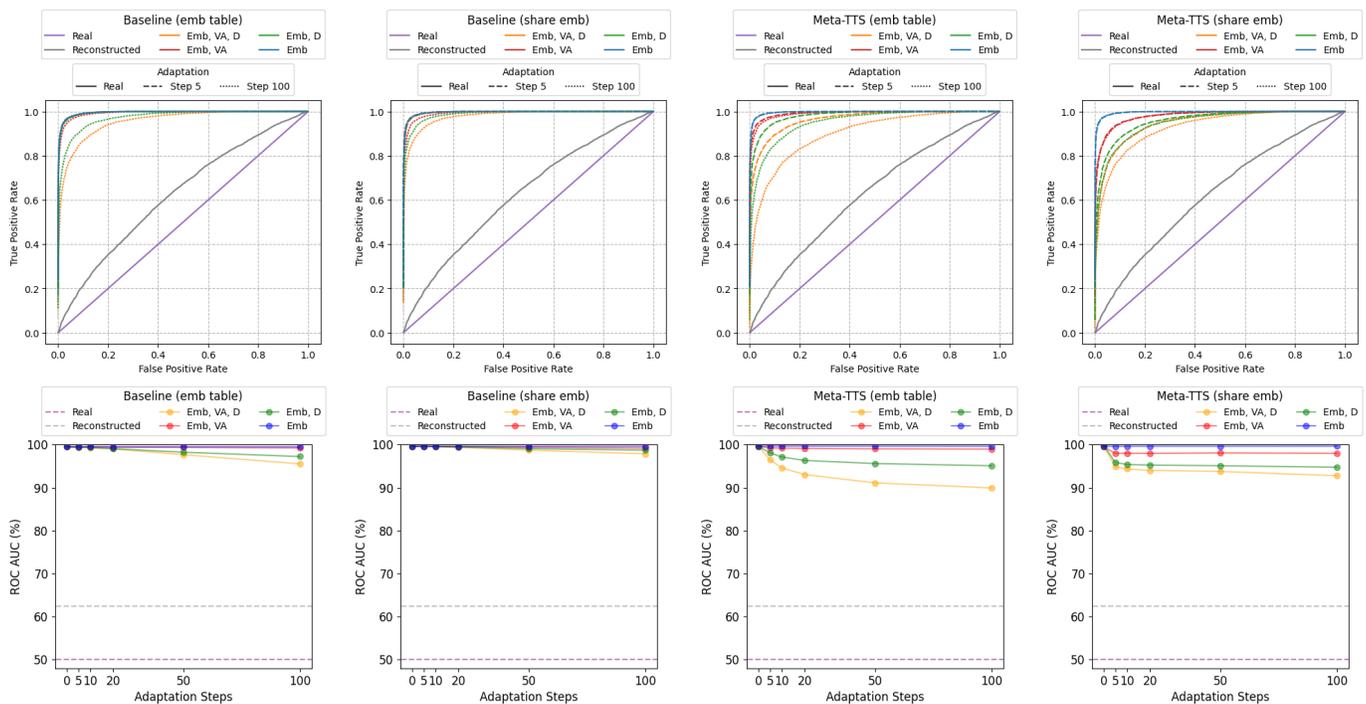

    \graphicspath{{VCTK/}}
    \centering
    \begin{subfigure}[]{0.24\linewidth}
        \includegraphics[width=\linewidth]{roc_base_emb.png}
    \end{subfigure}
    \hfill
    \begin{subfigure}[]{0.24\linewidth}
        \includegraphics[width=\linewidth]{roc_base_emb1.png}
    \end{subfigure}
    \hfill
    \centering
    \begin{subfigure}[]{0.24\linewidth}
        \includegraphics[width=\linewidth]{roc_meta_emb.png}
    \end{subfigure}
    \hfill
    \begin{subfigure}[]{0.24\linewidth}
        \includegraphics[width=\linewidth]{roc_meta_emb1.png}
    \end{subfigure}
    \\
    \centering
    \begin{subfigure}[]{0.24\linewidth}
        \includegraphics[width=\linewidth]{auc_base_emb.png}
    \end{subfigure}
    \hfill
    \begin{subfigure}[]{0.24\linewidth}
        \includegraphics[width=\linewidth]{auc_base_emb1.png}
    \end{subfigure}
    \hfill
    \centering
    \begin{subfigure}[]{0.24\linewidth}
        \includegraphics[width=\linewidth]{auc_meta_emb.png}
    \end{subfigure}
    \hfill
    \begin{subfigure}[]{0.24\linewidth}
        \includegraphics[width=\linewidth]{auc_meta_emb1.png}
    \end{subfigure}
    \caption{Synthesized speech detection on VCTK. \textbf{Top row:} ROC curves. \textbf{Bottom row:} ROC AUC.}
    \label{fig:synthesize-detection-VCTK}
\end{figure*}

After all, we evaluate how difficult it is to discern the synthesized utterances from the real ones.
The ROC curve of the detection results are in Figure~\ref{fig:synthesize-detection-LibriTTS} and~\ref{fig:synthesize-detection-VCTK}.
If the generated samples are indistinguishable from the real utterances, the ROC curve approaches the diagonal line, and the ROC AUC score should decrease to 50\%.
Compared with the baseline models, the observations are still the same: Meta-TTS fine-tuning the whole model (speaker embedding table + variance adaptor + decoder) performs the best.
However, the ROC AUC scores are far from 50\%, which means we can easily distinguish our generated utterances from the real utterances.
Our proposed Meta-TTS generates sounds of similar voices but distinguishable from the real utterances.
Nevertheless, this result does not violate our goal since we only aim to adapt the model efficiently but not generate undetectable fake speech.
Therefore, it is also good news that our proposed method can not deceive people for malicious purposes.

\subsection{Naturalness of synthesized samples}
\label{ssec:naturalness}

\begin{table*}[!h]
  \caption{Naturalness evaluated by MOS with 95\% confidence interval.}
  
  \centering
  \begin{tabular}{llcccc}
    \toprule
    \multirow{2}{*}{Approach}  & \multirow{2}{*}{Adaptation} & \multicolumn{2}{c}{LibriTTS} & \multicolumn{2}{c}{VCTK}\\
    \cmidrule(lr){3-4} \cmidrule(lr){5-6}
    &  & MOS & Estimated MOS & MOS & Estimated MOS\\
    \midrule
    \multicolumn{2}{l}{\textbf{Ground truth}}\\
    \addlinespace[0.2em]
    \multicolumn{2}{l}{Real utterances} 
        & $4.10 \pm 0.16$ & $3.70 \pm 0.03$
        & $4.04 \pm 0.11$ & $3.60 \pm 0.01$ \\
    \multicolumn{2}{l}{Reconstructed} 
        & $3.47 \pm 0.30$ & $3.55 \pm 0.03$
        & $3.38 \pm 0.15$ & $3.94 \pm 0.01$ \\
    \midrule
    \multicolumn{4}{l}{\textbf{With embedding table}}\\
    \addlinespace[0.2em]
    Baseline & 10 steps
        & $2.92 \pm 0.23$ & $\mathbf{3.66 \pm 0.02}$
        & $3.40 \pm 0.15$ & $\mathbf{3.79 \pm 0.01}$\\
    Meta-TTS & 10 steps
        & $\mathbf{2.95 \pm 0.19}$ & $3.34 \pm 0.03$
        & $\mathbf{3.44 \pm 0.14}$ & $3.55 \pm 0.01$\\
    \midrule
    \multicolumn{4}{l}{\textbf{With shared embedding}}\\
    \addlinespace[0.2em]
    Baseline & 10 steps
        & $\mathbf{3.67 \pm 0.22}$ & $\mathbf{3.82 \pm 0.02}$
        & $\mathbf{3.76 \pm 0.14}$ & $\mathbf{3.89 \pm 0.01}$\\
    Meta-TTS & 10 steps
        & $2.30 \pm 0.29$ & $2.67 \pm 0.04$
        & $3.09 \pm 0.14$ & $3.19 \pm 0.02$\\
    \bottomrule
  \end{tabular}
  \label{tab:mos}
\end{table*}

\begin{figure*}[!h]
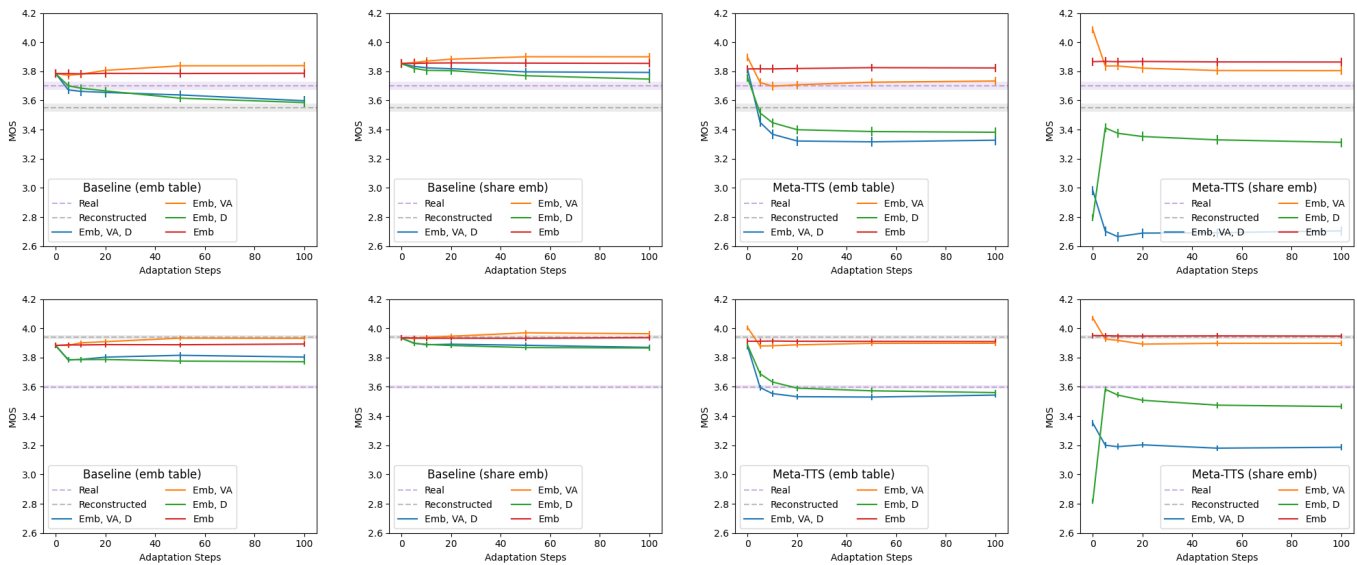

    \graphicspath{{LibriTTS/}}
    \centering
    \begin{subfigure}[]{0.24\linewidth}
        \includegraphics[width=\linewidth]{MOS_base_emb.png}
    \end{subfigure}
    \hfill
    \centering
    \begin{subfigure}[]{0.24\linewidth}
        \includegraphics[width=\linewidth]{MOS_base_emb1.png}
    \end{subfigure}
    \hfill
    \centering
    \begin{subfigure}[]{0.24\linewidth}
        \includegraphics[width=\linewidth]{MOS_meta_emb.png}
    \end{subfigure}
    \hfill
    \centering
    \begin{subfigure}[]{0.24\linewidth}
        \includegraphics[width=\linewidth]{MOS_meta_emb1.png}
    \end{subfigure}
    \\
    
    \graphicspath{{VCTK/}}
    \centering
    \begin{subfigure}[]{0.24\linewidth}
        \includegraphics[width=\linewidth]{MOS_base_emb.png}
    \end{subfigure}
    \hfill
    \centering
    \begin{subfigure}[]{0.24\linewidth}
        \includegraphics[width=\linewidth]{MOS_base_emb1.png}
    \end{subfigure}
    \hfill
    \centering
    \begin{subfigure}[]{0.24\linewidth}
        \includegraphics[width=\linewidth]{MOS_meta_emb.png}
    \end{subfigure}
    \hfill
    \centering
    \begin{subfigure}[]{0.24\linewidth}
        \includegraphics[width=\linewidth]{MOS_meta_emb1.png}
    \end{subfigure}
    \caption{Neural MOS estimation of naturalness on LibriTTS (top row) and VCTK (bottom row).}
    \label{fig:estimated-mos}
\end{figure*}

In this subsection, we measure the audio quality of the reconstructed utterances and the synthesized samples generated from the adapted models with human MOS ratings and neural MOS estimations.
Similar to SMOS in Section~\ref{ssec:similarity}, we only evaluate human MOS of fine-tuning the whole model with  10 adaptation steps due to the labor cost.
We show the results in Table~\ref{tab:mos} and Figure~\ref{fig:estimated-mos}.
In Table~\ref{tab:mos}, we could find out that with embedding table $\hat{E_S}$, Meta-TTS could give comparable naturalness to the baseline model, but shared embedding $e_S$ offers worse audio quality for Meta-TTS while the baseline model improves.
Predictably, sharing speaker embedding would harm the audio quality since the model is more vulnerable, as explained in Section~\ref{ssec:visualize}.
However, for the baseline model with $e_S$, since it cannot save any speaker information, the model probably becomes good at generating high-quality speech according to the input data without considering any speaker characteristics.
On the other hand, although other evaluation metrics show that Meta-TTS could improve speaker adaptation results by fine-tuning the decoder, the audio quality becomes worse, as shown in Figure~\ref{fig:estimated-mos}.
This result indicates that there is probably a trade-off between speaker similarity and audio quality for few-shot speaker adaptation tasks.



\section{Comparing with speaker-encoding baseline}
\label{sec:speaker-encoding-exp}
As stated earlier in the introduction section, there are mainly two general approaches to deal with the few-shot voice cloning tasks: speaker adaptation (with speaker embedding) and speaker encoding (with speaker encoder).
We have already compared the proposed meta version of the speaker adaptation approach with the standard version in Section~\ref{sec:exp-setup} and~\ref{sec:exp}.
In this section, we want to compare them with the speaker encoding approach.
Unlike tuning the speaker embedding, the speaker encoding approach encodes the speaker representations from the reference utterances without fine-tuning.
Since the speaker encoding approach does not require fine-tuning, it can provide the fastest voice cloning speed, but it can not further improve the cloning results according to the few-shot data, which is a trade-off.

In our experiments, we use GE2E model~\cite{wan2018generalized} as the speaker encoder, which is the same speaker verification model we used to extract $d$-vectors (Section~\ref{ssec:d-vector}).
Overall, we experiment under the three speaker encoder settings: jointly train with the TTS model (``scratch encoder''), pre-trained then fixed (``$d$-vector''), and pre-trained then jointly train with the TTS model (``pre-trained encoder'').
As mentioned in Section~\ref{ssec:d-vector}, the speaker encoder pre-training requires additional 3201 hours of audio data.

\paragraph{Models}
In this section, we will compare 7 settings:
\begin{itemize}[leftmargin=.6in]
    \item Real/Reconstructed utterances
    \item Meta-TTS and the speaker adaptation baseline \\(using speaker embedding table, fine-tuning speaker embedding + variance adaptor + decoder)
    \item 3 speaker encoding baselines (``scratch encoder'', ``$d$-vector'', ``pre-trained encoder'')
\end{itemize}

\paragraph{Training}
The training procedure is mostly the same as the speaker adaptation baseline, except we use the ground truth utterance as the reference audio and forward through the speaker encoder to get the speaker representation.

\paragraph{Inference}
During inference, we use the support sets' target utterances as the query sets' reference audios.
Since we are evaluating with 5-shot tasks, we would have 5 reference audios for each task.
We first encode these reference audios into vectors, then average them into a single speaker representation.

\subsection{Experimental results}
\begin{figure}[!h]
    \centering
    \begin{subfigure}[]{0.49\linewidth}
        \includegraphics[width=\linewidth]{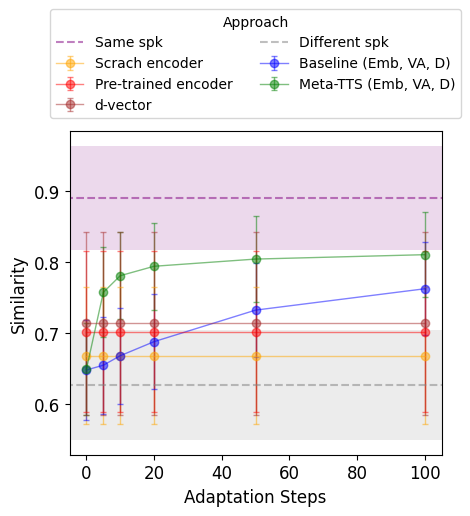}
        \caption{Similarity results on LibriTTS.}
    \end{subfigure}
    \hfill
    \begin{subfigure}[]{0.49\linewidth}
        \includegraphics[width=\linewidth]{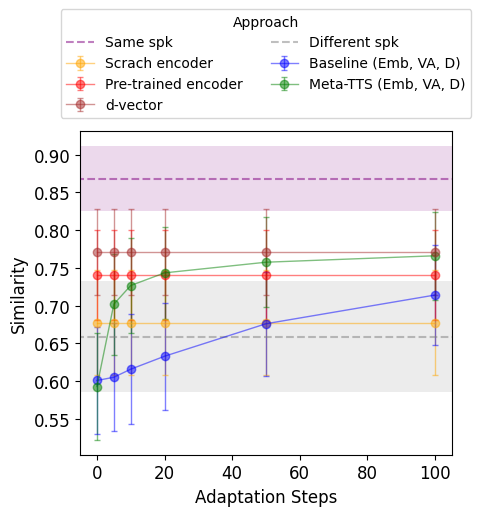}
        \caption{Similarity results on VCTK.}
    \end{subfigure}
    \caption{Speaker similarity results with the speaker-encoding baselines. The speaker encoding methods do not require fine-tuning, we plot them as horizontal lines with error-bars just for clearer comparison.}
    \label{fig:spk-similarity-encoder}
\end{figure}

\begin{figure}[!h]
    \centering
    \begin{subfigure}[]{0.49\linewidth}
        \includegraphics[width=\linewidth]{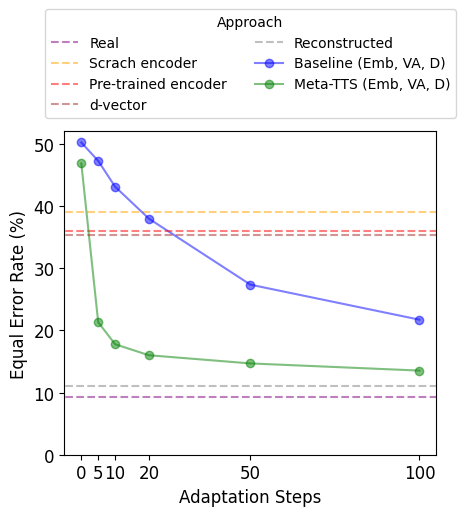}
    \end{subfigure}
    \hfill
    \begin{subfigure}[]{0.49\linewidth}
        \includegraphics[width=\linewidth]{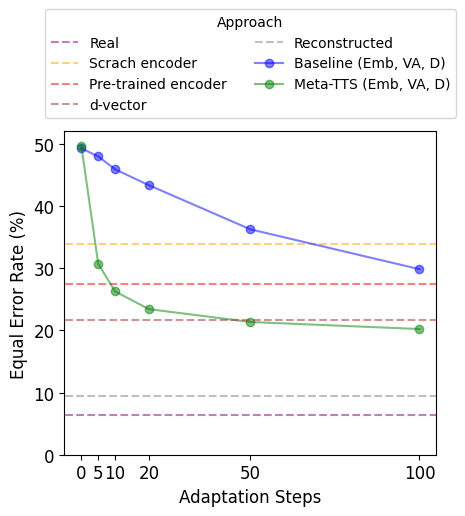}
    \end{subfigure}
    \\
    \centering
    \begin{subfigure}[]{0.49\linewidth}
        \includegraphics[width=\linewidth]{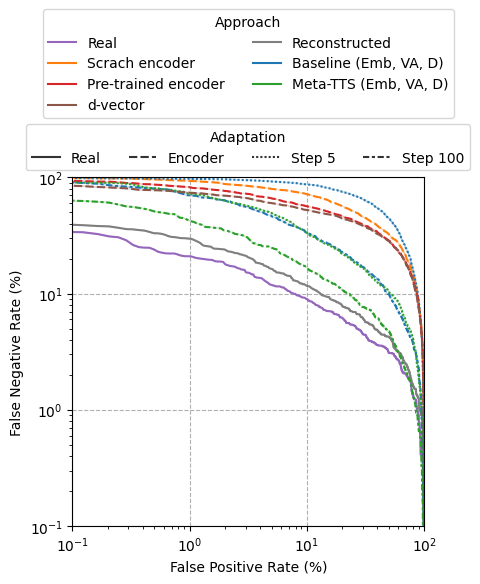}
    \end{subfigure}
    \hfill
    \begin{subfigure}[]{0.49\linewidth}
        \includegraphics[width=\linewidth]{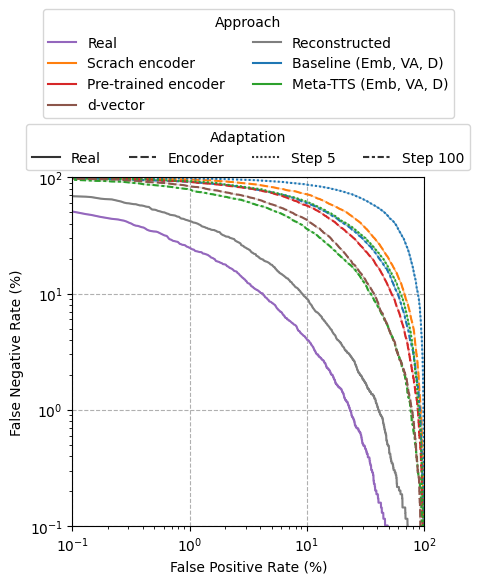}
    \end{subfigure}
    \caption{Speaker verification results. \textbf{Top:} EER. \textbf{Bottom:} DET curves. \textbf{Left:} LibriTTS. \textbf{Right:} VCTK.}
    \label{fig:sv-encoder}
\end{figure}

\begin{figure}[!h]
    \centering
    \begin{subfigure}[]{0.49\linewidth}
        \includegraphics[width=\linewidth]{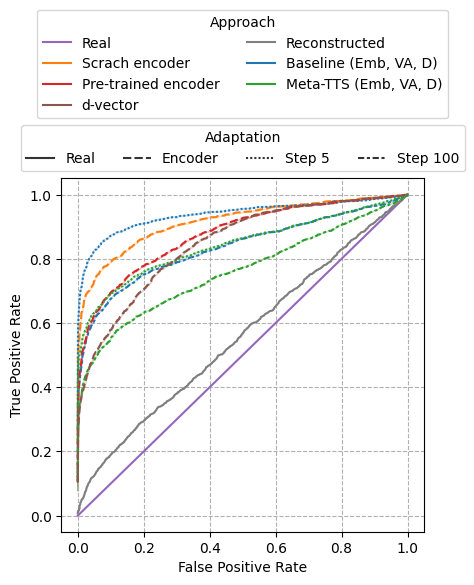}
    \end{subfigure}
    \hfill
    \begin{subfigure}[]{0.49\linewidth}
        \includegraphics[width=\linewidth]{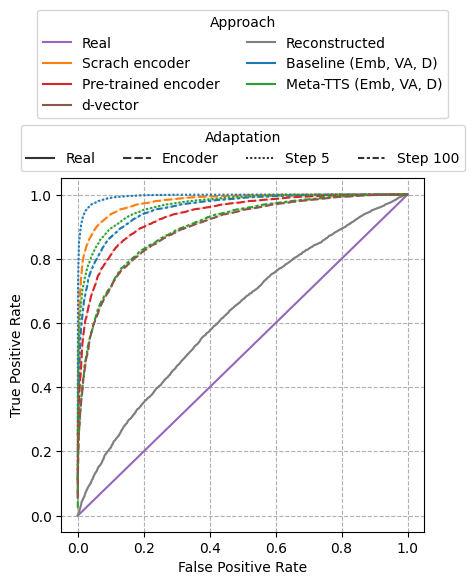}
    \end{subfigure}
    \\
    \centering
    \begin{subfigure}[]{0.49\linewidth}
        \includegraphics[width=\linewidth]{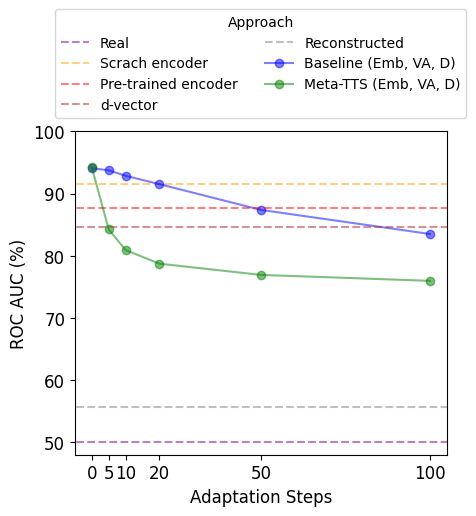}
    \end{subfigure}
    \hfill
    \begin{subfigure}[]{0.49\linewidth}
        \includegraphics[width=\linewidth]{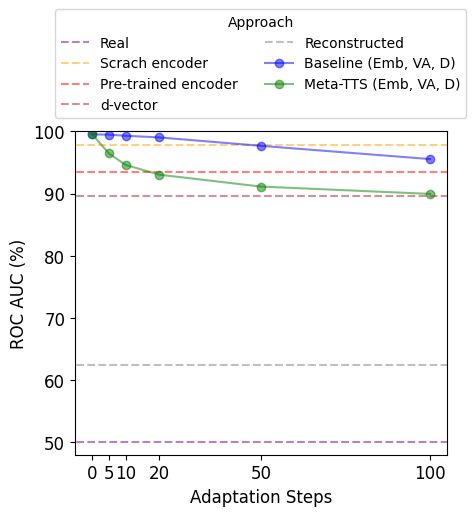}
    \end{subfigure}
    \caption{Synthesized speech detection results. \textbf{Top:} ROC curves. \textbf{Bottom:} ROC AUC. \textbf{Left:} LibriTTS. \textbf{Right:} VCTK.}
    \label{fig:synthesize-detection-encoder}
\end{figure}

The speaker similarity, speaker verification and synthetic speech detection results are in Figure~\ref{fig:spk-similarity-encoder},~\ref{fig:sv-encoder} and~\ref{fig:synthesize-detection-encoder} respectively.
The figures show that the best option for the speaker encoding methods is using $d$-vector (speaker encoder is pre-trained then fixed). Jointly train the speaker encoder with the TTS model leads to worse performance, whether the speaker encoder is trained from scratch or pre-trained.
This result shows that the speaker encoder can easily over-fit to the TTS training set, causing the generalization problem.

On the other hand, since the speaker encoder of the speaker encoding baselines causes more model parameters and can be pre-trained by additional large-scale datasets, it is not fair for Meta-TTS to compare with the speaker encoding baselines.
Nevertheless, Meta-TTS still performs comparably on VCTK and significantly outperforms the speaker encoding approach on LibriTTS within 100 adaptation steps and can further improve with more steps.
Therefore, Meta-TTS is the best voice cloning choice for both fast cloning with high voice similarity.
Moreover, it will be our future work to discover whether we can also apply meta-learning to the speaker encoding approach to minimize the generalization gap with few steps of the speaker encoder fine-tuning.

\section{Related works}
\cite{arik2018neural} uses DeepVoice3~\cite{ping2018deep} as the base model architecture.
It proposed the two basic voice cloning approaches, the speaker encoding method with a general speaker encoder and the speaker adaptation method by fine-tuning the speaker embedding with the model.
The experiments showed that, by fine-tuning the speaker embedding only, the speaker adaptation method performs comparably to the speaker encoding method (with 100k fine-tuning steps, which is around 8 hours of computation).
The speaker adaptation method performs even better by fine-tuning the whole model (adaptation would over-fit from hundreds to thousands of fine-tuning steps, depending on the number of fine-tuning samples).

SV2TTS~\cite{jia2018transfer} uses the speaker encoding method, where the speaker encoder is transferred from a pre-trained speaker verification model, and the TTS architecture is a Tacotron 2~\cite{shen2018natural}.
It showed a similar conclusion with \cite{arik2018neural} that with as less as 1 to 3 seconds (about 1 shot), adapting the speaker embedding only or using a speaker encoder would be a good choice without the need to consider over-fitting.
Nevertheless, adapting the whole model provides better performance when with more data (at least 3 shots).

SEA-TTS~\cite{chen2018sample} uses Wavenet~\cite{oord2016wavenet} as its basic architecture and compares the two voice cloning approaches.
As mentioned by \cite{arik2018neural} that adapting the whole model might result in over-fitting, SEA-TTS proposed two techniques to deal with it. First, SEA-TTS applies early stopping by splitting the few-shot support data into train and validation sets. Second, SEA-TTS would fine-tune the speaker embedding with 5k - 10k steps beforehand for better performance.

AdaSpeech~\cite{chen2021adaspeech} uses FastSpeech2 as its architecture with two model modifications for the speaker adaptation approach: acoustic condition modeling and conditional LayerNorm.
Then for speaker adaptation, the speaker embedding is jointly tuned with the conditional LayerNorm parameters for 2k steps.

In this paper, the speaker adaptation and speaker encoding baselines are the same as the proposed methods of the previous related works, while the TTS architecture is the only difference.
The speaker encoder of our speaker encoding baseline is even the same as SV2TTS.
However, unlike those related works, this paper mainly focuses on obtaining high-quality speaker adaptation results by fine-tuning within 100 steps.
Our paper shows that Meta-TTS could adapt efficiently, even when we only fine-tune the model for 10 to 20 steps.
Besides, we also compare our Meta-TTS with the speaker encoding method, which shows that even though the speaker encoder is pre-trained with additional 8371 speakers, our proposed Meta-TTS can still reach comparable or better results within 100 adaptation steps.

\section{Conclusion}
In this paper, we propose utilizing meta-learning for faster speaker adaptation on text-to-speech. We analyze the adaptation results from multiple perspectives and give out convincing conclusions that Meta-TTS could outperform the baseline multi-speaker TTS models significantly in both speaker similarity and adaptation speed.

\section{Acknowledgement}
We thank to National Center for High-performance Computing (NCHC) of National Applied Research 
Laboratories (NARLabs) in Taiwan for providing computational and storage resources.


%


\ifCLASSOPTIONcaptionsoff
  \newpage
\fi



\bibliographystyle{IEEEtran}
\bibliography{main}

\begin{IEEEbiography}
[{\includegraphics[width=1in,height=1.25in,clip,keepaspectratio]{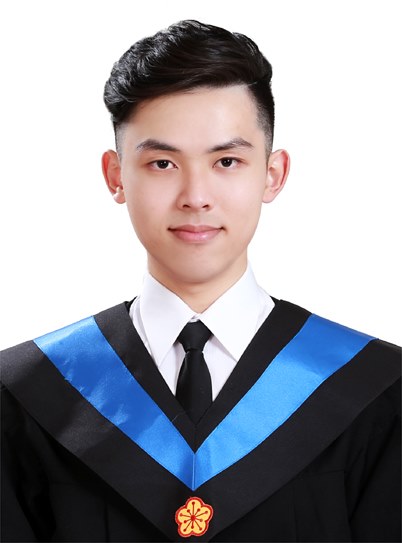} }]
{Sung-Feng Huang}
received the Bachelor degree from National Taiwan University (NTU) in 2017, and is now a P.h. D. student at the Graduate Institute of Communication Engineering (GICE) at National Taiwan University. He mainly works on speech recognition/separation/synthesis, spoken term detection, unsupervised/self-supervised/transfer/meta learning and machine learning techniques.
\end{IEEEbiography}

\begin{IEEEbiography}
[{\includegraphics[width=1in,height=1.25in,clip,keepaspectratio]{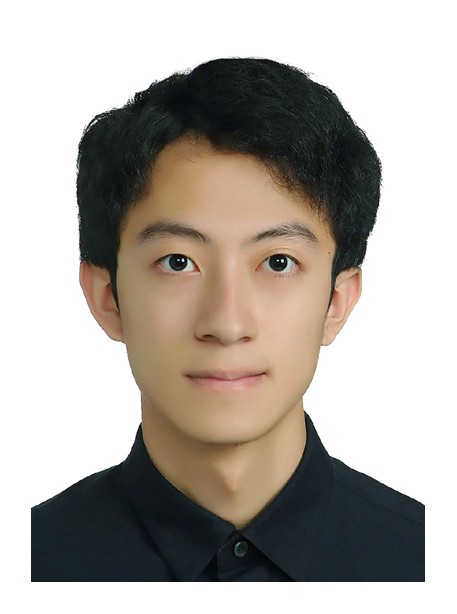} }]
{Chyi-Jiunn Lin}
is currently working toward a B.S degree in electrical engineering at National Taiwan University, from 2019. He is also an undergraduate research assistant at Speech Processing Lab, National Taiwan University. His research interests include spoken question answering, speech retrieval and speech synthesis.
\end{IEEEbiography}

\begin{IEEEbiography}
[{\includegraphics[width=1in,height=1.25in,clip,keepaspectratio]{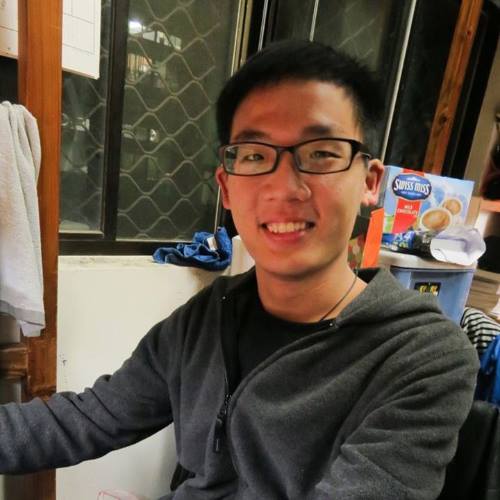} }]
{Da-Rong Liu}
received the Bachelor degree from National Taiwan University (NTU) in 2016, and is now a P.h. D. student at the Graduate Institute of Communication Engineering (GICE) at National Taiwan University. He mainly works on unsupervised learning, speech recognition and speech generation.
\end{IEEEbiography}

\begin{IEEEbiography}
[{\includegraphics[width=1in,height=1.25in,clip,keepaspectratio]{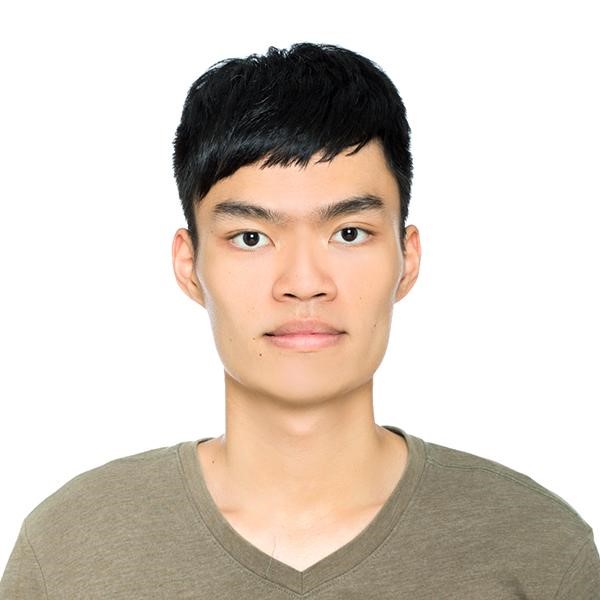} }]
{Yi-Chen Chen}
received the Bachelor degree from National Taiwan University (NTU) in 2017, and is now a P.h. D. student at the Graduate Institute of Communication Engineering (GICE) at National Taiwan University, working on self-supervised/semi-supervised/transfer learning and speech processing.
\end{IEEEbiography}

\begin{IEEEbiography}
[{\includegraphics[width=1in,height=1.25in,clip,keepaspectratio]{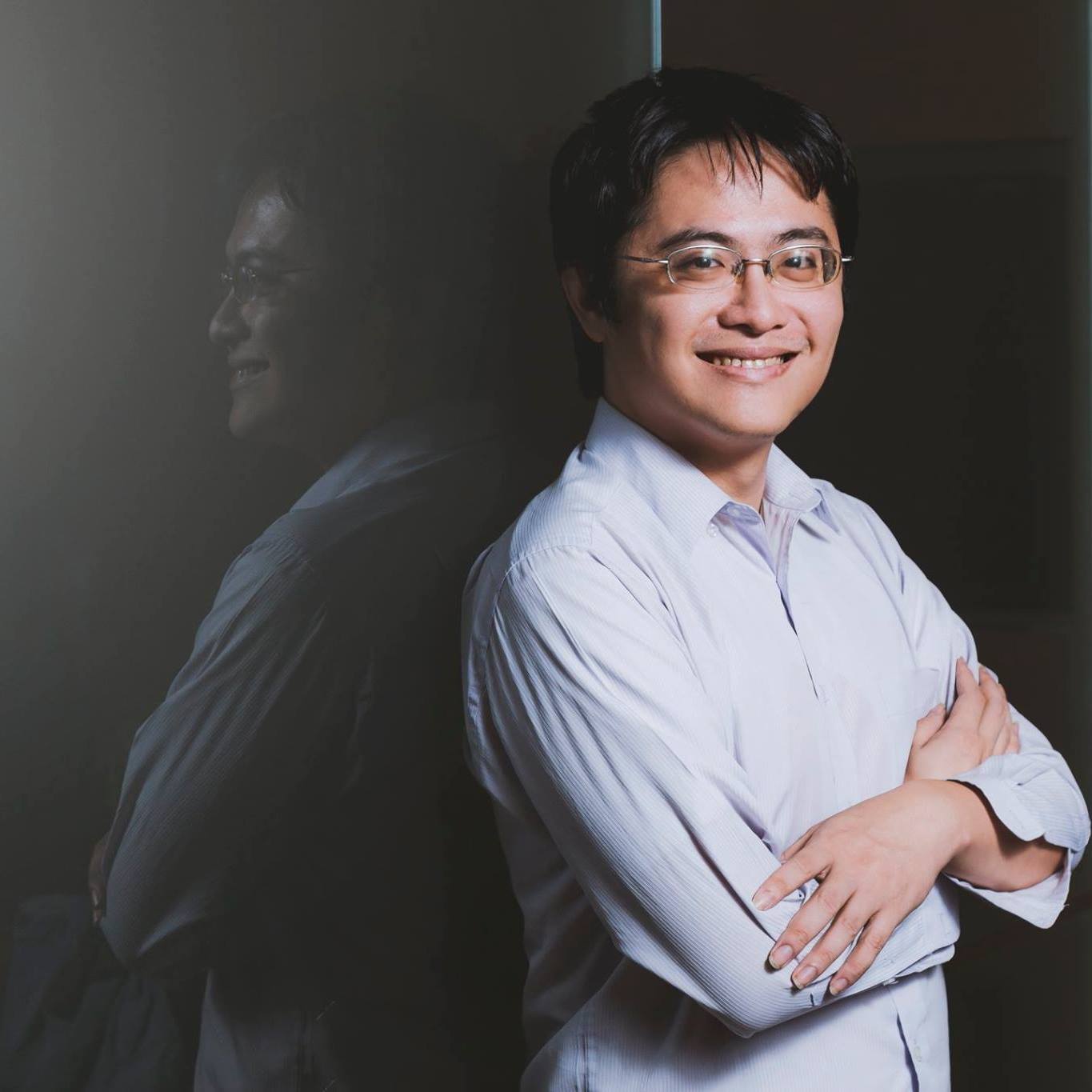} }]
{Hung-yi Lee} received M.S. and Ph.D. degrees from National Taiwan University in 2010 and 2012, respectively. From September 2012 to August 2013, he was a Postdoctoral Fellow with Research Center for Information Technology Innovation, Academia Sinica; from September 2013 to July 2014, he was a Visiting Scientist with the Spoken Language Systems Group, MIT Computer Science and Artificial Intelligence Laboratory. He is currently an Associate Professor in the Department of Electrical Engineering, National Taiwan University, with a joint appointment to the Department of Computer Science and Information Engineering. His research focuses on spoken language understanding, speech recognition, and machine learning.
\end{IEEEbiography}

\end{document}